\documentclass[aps,prd,eqsecnum,amsmath,amssymb,longbibliography,noeprint,10pt,twocolumn,floatfix,showkeys]{revtex4-2}

\usepackage{amsfonts}
\usepackage{mathtools}
\usepackage{graphicx}
\usepackage{dcolumn}
\usepackage{bm}
\usepackage{orcidlink}  
\usepackage{subcaption}
\usepackage[table,xcdraw,dvipsnames]{xcolor}
\usepackage{ragged2e}
\usepackage{soul}
\usepackage{url}
\usepackage{booktabs}
\usepackage{microtype}
\usepackage{hyperref}

\definecolor{blue-violet}{rgb}{0.7, 0.2, 0.8}

\sethlcolor{yellow}

\preprint{AIP/123-QED}

\allowdisplaybreaks

\newcommand{\dd}{\mathrm{d}}
\newcommand{\Msun}{M_\odot}
\newcommand{\Mc}{\mathcal{M}}

\newcommand{\Lpop}{\Lambda_{\rm pop}}

\newcommand{\Agwb}{A_{\rm gwb}}

\newcommand{\data}{d}
\newcommand{\cat}{C}

\begin{document}

\title{Hierarchical Inference of the Supermassive Black Hole Binary Merger Rates from Joint Searches using Pulsar Timing Arrays}

\author{Sharon Mary Tomson \orcidlink{0000-0001-7603-1637}}
 \email{sharon.mary.tomson@aei.mpg.de}
\author{Rutger van Haasteren \orcidlink{0000-0002-6428-2620}}
\author{Boris Goncharov \orcidlink{0000-0003-3189-5807}}%
\affiliation{ 
Max Planck Institute for Gravitational Physics (Albert Einstein Institute), 30167 Hannover, Germany\\
Leibniz Universität Hannover, 30167 Hannover, Germany
}%
\date{\today}

\begin{abstract}
Gravitational-wave searches do more than identify individual sources - they provide a way to infer the underlying astrophysical populations that produce them. Currently, Pulsar Timing Arrays (PTAs) constrain the supermassive black-hole binary (SMBHB) merger-rate density through both the stochastic gravitational-wave background (SGWB) and searches for individual SMBHB signals. The latter implies converting source upper limits into rate upper limits using detection efficiencies estimated empirically. We instead develop a hierarchical Bayesian framework that places the SMBHB merger-rate density model directly inside the PTA likelihood. The unknown catalog of individual SMBHB merger signals is modeled as a Poisson point process on source-parameter space, so that the number of merger signals is inferred from the data rather than imposed by a detection threshold. We describe two computational routes: a catalog-marginal likelihood based on analyses with fixed numbers of candidate sources, and an explicit transdimensional sampling approach that jointly samples population hyperparameters, latent merger signals, and noise. We validate the method with a toy merger-only population model and then apply it to an astrophysical SMBHB merger-rate density model that jointly predicts the SGWB amplitude and the expected merger-catalog size. In simulations, an individual merger signal adds complementary information to the SGWB and can tighten constraints on the merger-rate population.

\end{abstract}

\keywords{gravitational waves --- 
pulsars: general --- methods: data analysis}

\maketitle

\section{Introduction}
\label{sec:introduction}

Pulsar timing arrays (PTAs)~\cite{FosterBacker1990} transform an ensemble of precisely timed millisecond pulsars into a Galactic-scale detector for nanohertz gravitational waves. 
The idea of a PTA, first proposed in the late 1970s~\cite{Sazhin1978,Detweiler1979}, has recently entered an exciting observational era. 
Multiple PTA collaborations have now reported evidence for Hellings-Downs correlations~\cite{HellingsDowns1983} of a low-frequency stochastic gravitational-wave background (SGWB)~\cite{NG_15_GWB,EPTA_DR2_GWB,PPTA_DR3_GWB,CPTA_DR1_GWB,MT_DR1_GWB,IPTA_DR2_GWB,YuAllen2025}.

If the expected origin of the SGWB is confirmed~\cite{GoncharovSato-Polito2026}, gravitational waves in PTA data will reveal information about the astrophysical populations that generate them. A stochastic background sourced by SMBHBs encodes the integrated contribution of binaries across cosmic time, and therefore carries information about galaxy mergers, black-hole--host-galaxy relations, binary hardening mechanisms, environmental coupling, eccentricity, and the timescale between galaxy pairing and black-hole coalescence \cite{SesanaVecchio2008,Sesana2013,MiddletonDelPozzo2016,ChenSesana2019}. The central challenge is to extract this population information in a way that is statistically consistent with the source searches.

Current PTA constraints on SMBHB populations are obtained through three
related routes. First, measurements of the SGWB can be interpreted using
astrophysical SMBHB population models to constrain the population properties
that determine the background
\cite{MiddletonDelPozzo2016,ChenSesana2019,NG_15_HOLODECK,
GoncharovSardana2025}. In simpler implementations, the population is mapped
to a small number of summary properties of the SGWB, such as its amplitude
and spectral shape. Because these quantities describe the integrated
contribution of the SMBHB population, different combinations of merger rate,
chirp-mass distribution, redshift evolution, and binary astrophysics can
produce similar backgrounds. More detailed approaches retain substantially more of the frequency-dependent information in the SGWB. \citet{TaylorSimon2017} used population-synthesis simulations and Gaussian-process emulation to predict
astrophysically motivated SGWB spectra, allowing deviations from a simple
power law produced by stellar environmental coupling and orbital eccentricity to be incorporated directly into Bayesian PTA inference. More recently, \citet{LaalTaylor2025} developed a normalizing-flow emulator for the multivariate distribution of SGWB spectra, retaining correlations between frequency bins and non-Gaussian features of
the population predictions. Second, deterministic searches constrain individually resolvable continuous gravitational waves from SMBHBs residing in the PTA band \cite{ZhaoChen2025,AgazieAnumarlapudi2023,EPTACollaborationInPTACollaboration2024}. Third, more recent approaches jointly model the
SGWB and individual continuous-wave sources through shared population-level parameters \cite{GoncharovSato-Polito2026}.

Contemporary CGW analyses report the merger-rate density of SMBHBs, inferring it indirectly from the local number density of binaries evaluated as
\begin{equation}
    n_{\rm c}^{95}
    =
    \frac{-\ln(1-C)}{V_{\rm c}},
\end{equation}
where \(C=0.95\) is the confidence level used to define the upper limit, and \(V_{\rm c}\) is the comoving volume out to the distance limit.
It is based on the
expected number of detectable binaries, 
$N_{\rm exp}^{\rm det}=n_{\rm c} V_{\rm c}$.
For zero detections, assuming Poisson statistics, $P(0\mid N_{\rm exp}^{\rm det})=\exp[-N_{\rm exp}^{\rm det}]$. 
To convert a CGW number-density constraint to a merger-rate constraint, one uses the residence time of binaries
in the PTA frequency band. If \(R\) is the local merger-rate density and \(\Delta T(f_i)\) is the time a binary spends in frequency bin \(i\), then the expected number of detected CGW sources can be written as
\begin{equation}
    N_{\rm exp}^{\rm det}
    =
    R
    \sum_i
    \epsilon_{\rm det}\,
    V_i\,
    \Delta T(f_i),
\end{equation}
where \(V_i\) is the sensitive volume in the \(i\)-th frequency bin and \(\epsilon_{\rm det}\) is the detection efficiency. In the PPTA DR3 CGW analysis, ~\citet{ZhaoChen2025}, following ~\citet{ZhuHobbs2014}, this efficiency is set to \(\epsilon_{\rm det}=0.95\). The corresponding 95\% upper limit on the merger rate is
\begin{equation}
    R_{95}
    =
    \frac{-\ln(1-0.95)}
    {
    \sum_i
    \epsilon_{\rm det}\,
    V_i\,
    \Delta T(f_i)
    } .
\end{equation}
Thus, constrains on merger-rate density from CGW searches require an additional detection efficiency calculation. 

Although this construction provides a way to translate CGW search results into constraints on the SMBHB merger rate, the merger rate is not inferred directly from the PTA data. Instead, the analysis first constrains the number of binaries residing in the PTA band and then converts this number-density constraint into a merger-rate constraint using the binary residence time and the sensitive volume. This separation introduces several limitations. First, current CGW-based merger-rate calculations typically provide an upper limit rather than a posterior on the underlying merger-rate population. Second, it is indirect, as most of the metric perturbations produced by gravitational waves in the PTA band at Earth do not significantly evolve in frequency over decades of observations, and thus do not exhibit SMBHB mergers. Third, the resulting rate constraint depends on how the sensitivity of the CGW search is characterized. A threshold-based search requires a detection
threshold defined relative to a background, or null, distribution of the search statistic. In PTA analyses this distribution depends on the assumed pulsar-noise model, the timing-model fit, and any common processes included in the likelihood. Recent work has emphasized that PTA analyses cannot generate fully empirical samples from the true null hypothesis
without making modeling assumptions, so background
estimates based on scrambling or related procedures are
not completely model-independent~\cite{vanHaasteren2025}. The corresponding detection efficiency must then be calibrated, typically through signal injections, and consequently depends on the assumed source population, waveform model, noise realization, and recovery criterion. These large injection studies to derive the threshold can be computationally expensive.

These complications motivate inferring the SMBHB merger-rate population directly within the PTA likelihood rather than converting deterministic-search outputs into rate constraints as a separate post-processing step. In this work, we model the unknown catalog of SMBHB merger signals as a Poisson point process (PPP) whose intensity is determined by the astrophysical merger-rate population. The number of merger signals, their source parameters, the population hyperparameters, and the PTA noise parameters can therefore be inferred within a common hierarchical model. Weak merger signals and nondetections contribute through the likelihood without requiring a separate
detection threshold or detection-efficiency correction.

This formulation also provides a natural connection between individual merger signals and the SGWB. The unresolved SGWB and individual SMBHB mergers are different observables of the same underlying astrophysical population. Changes in the merger-rate normalization, chirp-mass distribution, or redshift evolution affect both the accumulated background and the expected number and properties of individual merger signals. Consequently, the two observables can provide complementary constraints on the same merger-rate population. For simplicity, we approximate SMBHB mergers by only their dominant gravitational wave memory contribution (bursts with memory) to reduce the computational cost \cite{vanHaasterenLevin2010}. However, the formalism we present can be extended to a full signal from SMBHB mergers, including also the inspiral, the merger, and the ringdown~\cite{TomsonGoncharov2026}.

Section~\ref{sec:population_to_data_model} defines the hierarchical population model, including the source-frame merger-rate density, its mapping to PTA observables, the Poisson point-process catalog prior, and the PTA likelihood implementations. Section~\ref{sec:implementation} describes the simulated PTA datasets and the inference settings used for the catalog-marginalized and transdimensional analyses. Section~\ref{sec:results_toy} validates the unknown-catalog inference on a toy merger-only population, and Sec.~\ref{sec:results_joint} applies the framework to a joint SGWB+merger population analysis. Section~\ref{sec:discussion} discusses limitations and future extensions, and Sec.~\ref{sec:conclusion} summarizes the main results.

\section{Hierarchical population model}
\label{sec:population_to_data_model}

Hierarchical Bayesian inference is standard in gravitational-wave astronomy: event-level likelihoods constrain source parameters, while a population model describes the distribution from which sources are drawn. 
At the lowest level, the data constrain source or signal parameters. At the population level, those parameters are assumed to be drawn from a distribution
controlled by hyperparameters. Schematically, if $s$ denotes
source-level parameters and \(\Lambda\) denotes population hyperparameters, a
single-event hierarchical likelihood marginalized over event parameters has the form
\begin{equation}
    p(d\mid\Lambda)
    =
    \int
    p(d\mid s)\,
    p(s\mid\Lambda)\,
    \dd s ,
\end{equation}
where $d$ is data. 
When the event count is also informative, the population is naturally modeled
as an inhomogeneous Poisson process. Let
\(\mathcal{R}(s\mid\Lambda)\) be the differential event-rate density in
source-parameter space. This may be decomposed into an overall rate and a
population shape, for example
\(\mathcal{R}(s\mid\Lambda)=\dot n_0\,p(s\mid\Lambda)\). For a
threshold-selected catalog of detected events, selection effects enter through
the detection probability \(p_{\rm det}(s)\), giving the expected number
of detected events
\begin{equation}
    \lambda_{\rm det}(\Lambda)
    =
    T_{\rm obs}
    \int
    p_{\rm det}(s)\,
    \mathcal{R}(s\mid\Lambda)\,
    \dd s .
\end{equation}
The corresponding population likelihood can be written schematically as
\begin{equation}
    p(\{d_j\}\mid\Lambda)
    \propto
    e^{-\lambda_{\rm det}(\Lambda)}
    \prod_{j=1}^{N_{\rm obs}}
    \int
    p(d_j\mid s)\,
    \mathcal{R}(s \mid\Lambda)\,
    \dd s .
\end{equation}
This is the structure used in ground-based gravitational-wave population
analyses to infer compact-binary merger rates and population hyperparameters
while accounting for selection effects
\cite{ThraneTalbot2019,MandelFarr2019,LVK_GWTC3_Merger,TaylorGerosa2018,VitaleGerosa2022}. 

Recent PTA analyses have also used hierarchical methods to model ensemble-level properties of pulsars, for example by describing the distribution of intrinsic pulsar red-noise amplitudes and spectral indices with shared hyperparameters \cite{vanHaasteren2024,GoncharovSardana2025_a}. These applications differ from
the merger-rate problem considered here, but they illustrate the same statistical point that PTA data can contain population-level information that should be modeled hierarchically. The PTA merger problem considered here is different from a threshold-selected catalog of detected events. We do not first identify a set of detected merger signals and then correct the population likelihood using a separate detection probability \(p_{\rm det}(s)\). Instead, the unknown merger catalog is treated as a latent variable inside the PTA likelihood. The probability of zero, one, or more merger signals is inferred directly from the data through the Poisson point-process prior and the PTA likelihood. In this formulation, the analog of selection is encoded by the likelihood itself where the population models that would produce signals inconsistent with the data receive low likelihood, while weak signals and nondetections contribute through posterior support.

\begin{figure*}
    \centering
    \includegraphics[width=0.9\linewidth]{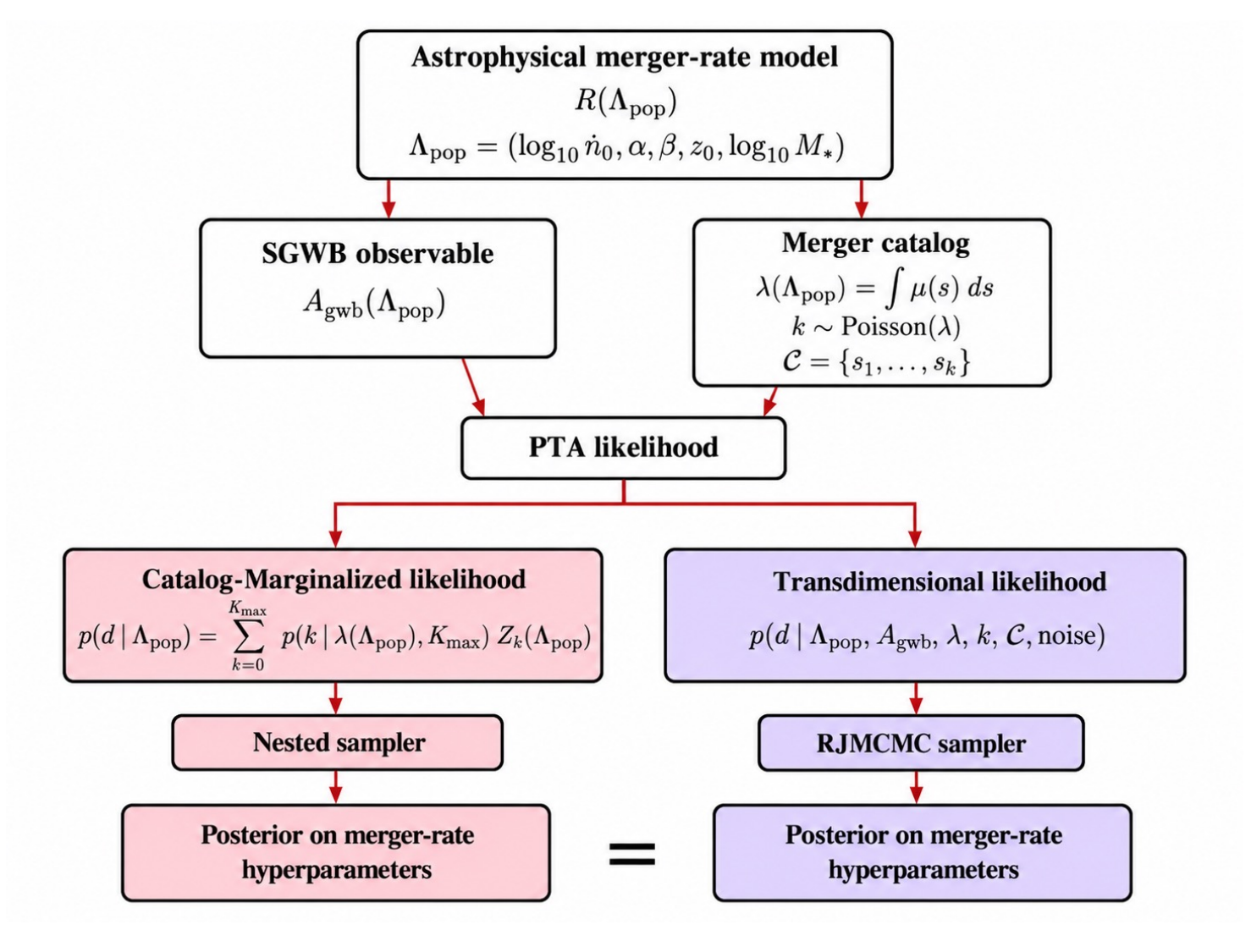}

    \caption{\justifying
    Schematic of the hierarchical population model. A source-frame merger-rate density
    model defines the population hyperparameters. This rate is mapped to
    PTA-frame observables: the expected number and distribution of merger
    signals, and the SGWB.
    }
    \label{fig:population_to_data_flow}
\end{figure*}

We construct a hierarchical SMBHB merger-rate population model for PTAs, summarized in Figure~\ref{fig:population_to_data_flow}. 
The hierarchy has four conceptual steps:
\begin{enumerate}
    \item A merger-rate density $R$ specifies the astrophysical
    population through hyperparameters $\Lpop$.
    \item This density is mapped to two PTA-frame observables: a merger-catalog intensity $\mu(s\mid\Lambda_{\rm pop})$, which gives the expected number density of merger signals in source parameter space, and the amplitude of the unresolved SGWB $A_{\rm{gwb}}$.
    \item The unknown catalog of individual SMBHB mergers is
    modeled as a Poisson point process.
    \item The PTA likelihood compares the data to a model containing
    the deterministic merger catalog, the SGWB, and the pulsar
    noise models.
\end{enumerate}

For each merger \(i\), we distinguish the source parameters that are controlled by the astrophysical population model from the additional parameters required to construct its PTA waveform. We denote the population-dependent source parameters by $s_i =\left(\log_{10}\Mc_i,\,z_i\right)$ where \(\Mc_i\) is the source-frame chirp mass and \(z_i\) is the redshift. For the assumed cosmology, the luminosity distance \(D_{{\rm L},i}\) is determined by \(z_i\). The remaining source and waveform parameters are denoted collectively by
\(\vartheta_i\). These include the observer-frame merger time \(t_{0,i}\),
sky location $\theta,\phi$, polarization angle $\psi$, and any other waveform parameters whose
priors are independent of \(\Lpop\). In the waveform implementation, it is
convenient to evaluate the population-dependent part of the source model
using \((D_{{\rm L},i},\log_{10}\Mc_i)\), with \(t_{0,i}\) included
separately among the waveform parameters. PTA noise parameters are denoted
by \(n\).

\subsection{Source-frame merger-rate model}
\label{subsec:source_frame_rate_model}

We use the phenomenological SMBHB merger-rate model introduced by
\citet{MiddletonDelPozzo2016}. The model was designed to capture the broad
mass and redshift dependence expected for cosmological SMBHB merger
populations without assuming a particular galaxy- or black-hole-assembly
history. Its chirp-mass dependence is Schechter-like, consisting of a
power law with an exponential suppression above a characteristic mass
\(M_*\), while the redshift dependence is described by a power law in
\(1+z\) with an exponential suppression on the characteristic scale \(z_0\).
This form is sufficiently flexible to reproduce the qualitative behavior of
merger rates obtained from merger-tree models and cosmological simulations
\cite{MiddletonDelPozzo2016}.

We define the corresponding source-frame merger-rate density as
\[
    R(\log_{10}\Mc,z\mid\Lpop)
    [{\rm Mpc}^{-3}\,{\rm Gyr}^{-1}\,
    {\log_{10}M_\odot}^{-1}],
\]
that is, the merger rate per comoving volume, per source-frame time, and per
logarithmic chirp-mass interval. The model is
\begin{equation}
\begin{split}
\label{eq:Rsrc_pheno}
    R(\log_{10}\Mc,z\mid\Lpop)
    &=
    \dot n_0
    \left(\frac{\Mc}{10^7\Msun}\right)^{-\alpha}
    \exp\!\left[-\frac{\Mc}{M_*}\right]
     \\
    &\quad \times
    (1+z)^{\beta}
    \exp\!\left[-\frac{z}{z_0}\right] .
\end{split}
\end{equation}

The population hyperparameters are
\begin{equation}
\label{eq:population_hyperparameters}
    \Lpop =
    \left(\log_{10}\dot n_0,\alpha,\beta,z_0,\log_{10}M_*\right).
\end{equation}
Here \(\dot n_0\) $[{\rm Mpc}^{-3}\,{\rm Gyr}^{-1}\,{\log_{10}M_\odot}^{-1}]$ is the source-frame merger-rate density normalization having the same units as $R$. The
parameters \(\alpha\) and \(M_*\) control the chirp-mass dependence and
high-mass turnover, while \(\beta\) and \(z_0\) control the redshift evolution and high-redshift evolution turnover. 
In the present proof-of-concept model, the mass-ratio dependence is not sampled; we assume equal-mass SMBHBs.

This differs in notation from Eq.~(2) of
\citet{MiddletonDelPozzo2016}, where the corresponding distribution is
written per unit redshift and therefore contains an additional factor
\(\lvert dt_{\rm R}/dz\rvert\). Here we define \(R\) directly per unit
source-frame time; the time--redshift conversion is introduced explicitly
when mapping the rate density to PTA observables below. 

\subsection{From astrophysical rate density to PTA observables}
\label{subsec:rate_to_pta_observables}

The same source-frame merger-rate density determines two PTA observables:
individual SMBHB merger signals and the unresolved SGWB. For the
individual-merger channel, the merger-rate population determines both the
properties of each deterministic merger signal and the expected number and
distribution of such signals in the PTA dataset. For the SGWB channel, the
same population determines the accumulated unresolved inspiral signal from
binaries radiating in the PTA band. These two observables are therefore
generated by the same underlying merger-rate density \(R\), but through
physically distinct mappings.

\subsubsection{Individual SMBHB Mergers : signal model}
\label{merger_memory_signal}

PTAs do not observe SMBHB mergers in the same way as ground-based detectors.
The high-frequency merger and ringdown signals occur on timescales
much shorter than the cadence of pulsar timing observations. What remains
as an observable in PTA data is the non-oscillatory gravitational-wave memory -
a permanent change in the metric that appears as a step function in strain and,
after integration, as a ramp in timing residuals \cite{Favata2010,vanHaasterenLevin2010}. This ``memory burst''
approximation is well motivated when the memory growth time is shorter than
the sampling interval of the PTA data \cite{vanHaasterenLevin2010}.

In the present work we use this standard burst-with-memory approximation for
the deterministic merger signal. The Earth-term strain is modeled as a
step,
\begin{equation}
\label{eq:memory_step_strain}
    h_+(t)
    =
    h_{\rm mem}\,\Theta(t-t_0)
\end{equation}
where \(h_{\rm mem}\) is the memory amplitude and \(t_0\) is the merger time. For one pulsar, the induced
timing residual is
\begin{equation}
\label{eq:memory_ramp_residual}
    r_i(t)
    =
    F_i(\hat{\Omega},\psi)\,
    h_{\rm mem}\,
    (t-t_0)\Theta(t-t_0)
\end{equation}
where \(F_i\) is the pulsar antenna response, depending on the pulsar sky
position, the GW propagation direction \(\hat{\Omega}\), and the polarization
angle \(\psi\). 

For a binary merger, we relate the step amplitude to the source
parameters using the approximate accumulated memory amplitude
\begin{equation}
\label{eq:hmem_amplitude}
    h_{\rm mem}
    =
    \frac{G}{24 c^2}
    \frac{\eta M}{D_{\rm{c}}}
    \sin^2\iota\,
    \left(17+\cos^2\iota\right)
    \left[
    \frac{\Delta E_{\rm rad}}{\eta M c^2}
    \right],
\end{equation}
where \(M\) is the total mass, \(\eta\) is the symmetric mass ratio, \(D_{\rm{c}}\) is the comoving distance to the source, \(\iota\) is the inclination, and
\(\Delta E_{\rm rad}\) is the radiated energy \cite{AggarwalArzoumanian2020}. In the current astrophysical implementation, we assume an equal-mass binary with
\(\eta=0.25\), fix the inclination to \(\iota=\pi/3\), and take $\Delta E_{\rm rad}/{\eta M c^2}=1-\sqrt{8}/3$. Using \(\Mc=\eta^{3/5}M\), so that
\(\eta M=\eta^{2/5}\Mc\), and
$D_{\rm c}=D_{\rm L}/{1+z}$, Eq.~\eqref{eq:hmem_amplitude} reduces to
\begin{equation}
\label{eq:hmem_reduced}
    h_{\rm mem}
    \simeq
    8.47\times10^{-16}
    (1+z)
    \left(\frac{\Mc}{10^9\,M_\odot}\right)
    \left(\frac{1\,{\rm Gpc}}{D_{\rm L}}\right).
\end{equation}

This signal model is intentionally simple. It captures the leading PTA
signature of memory as a coherent ramp across the array, but it does not model
the gradual build-up of memory during the late inspiral and merger, nor the
oscillatory inspiral--merger--ringdown part of the signal. More complete
waveform models now include both the oscillatory SMBHB waveform and the nonlinear
memory contribution, and show that the step-function approximation can bias
the recovered amplitude and source parameters for sufficiently strong nearby
systems~\cite{TomsonGoncharovHaasteren2026}. In this paper, we use the step-memory model
as a controlled first implementation of the hierarchical population model. It may be of interest to replace it with the full merger waveform in future work.

\subsubsection{Individual SMBHB mergers: catalog intensity and poisson point process}
\label{subsubsec:merger_catalog_model}

For the present proof-of-concept analysis, we restrict the explicitly
modeled merger catalog to a nearby, high-mass region. We introduce the
fixed source hyperparameters
\(\Mc_{\min}\), \(\Mc_{\max}\),
\(D_{{\rm L},\min}\), and \(D_{{\rm L},\max}\).
Because the population intensity is written in the coordinates
\(s=(\log_{10}\Mc,z)\), the corresponding catalog domain is
$\mathcal{D}_{\rm m}= \Mc_{\min}\leq\Mc\leq\Mc_{\max},  z(D_{{\rm L},\min})\leq z \leq
    z(D_{{\rm L},\max})$.

Merger events outside \(\mathcal{D}_{\rm m}\), particularly lower-amplitude
or more distant systems, are not included as explicit deterministic
sources in the present analysis. A population of unresolved merger-memory
signals can in principle generate a stochastic gravitational-wave memory
background~\cite{ZhaoCao2022}. This contribution is expected to be
subdominant to the SGWB from a population of inspiralling SMBHBs and hence is neglected here. The bounds defining \(\mathcal{D}_{\rm m}\) apply only to the explicitly modeled merger catalog and do not restrict the SGWB integration domain.

Within this catalog domain, the source-frame merger-rate density is converted
to an observer-frame catalog intensity by accounting for cosmological volume
and time dilation. Integrated over an observing duration \(T_{\rm obs}\),
the intensity in the population coordinates
\(s=(\log_{10}\Mc,z)\) is
\begin{equation}
\label{eq:mu_merger_simple}
\begin{split}
    \mu(\log_{10}\Mc,z\mid\Lpop)
    =
    &\,10^{-9} T_{\rm obs}\,
    R(\log_{10}\Mc,z\mid\Lpop)
    \\
    &\times
    \frac{\dd V_c}{\dd z}
    \frac{1}{1+z}.
\end{split}
\end{equation}
It is defined such that
\(\mu(s\mid\Lpop)\,\dd s\) is the expected number of merger signals
within the infinitesimal source-parameter volume \(\dd s\) during the
observing interval. Here \(T_{\rm obs}\) is expressed in years, and the
factor \(10^{-9}\,{\rm Gyr}\,{\rm yr}^{-1}\) converts the source-frame
rate from \({\rm Gyr}^{-1}\) to \({\rm yr}^{-1}\). The factor
\((1+z)^{-1}\) accounts for cosmological time dilation. Consequently,
$[\mu(\log_{10}\Mc,z\mid\Lpop)]=[{\log_{10}M_\odot}^{-1}]$.

The expected number of merger signals in the PTA dataset is therefore
\begin{equation}
\label{eq:lambda_merger_simple}
    \lambda(\Lpop)
    =
    \int_{\mathcal{D}_{\rm m}}
    \mu(s\mid\Lpop)\,\dd s .
\end{equation}
Thus \(\lambda(\Lpop)\) is dimensionless and is derived from the
source-frame merger-rate density rather than sampled as an independent
population parameter.

Conditional on a merger occurring within the explicit catalog domain,
the normalized density of its population-dependent source parameters is
\begin{equation}
\label{eq:normalized_source_density}
    p(s\mid\Lpop)
    =
    \frac{
        \mu(s\mid\Lpop)
    }{
        \lambda(\Lpop)
    },
    \qquad
    s\in\mathcal{D}_{\rm m}.
\end{equation}

The quantities \(\lambda(\Lpop)\) and \(p(s\mid\Lpop)\) describe the
expected catalog size and source-parameter distribution, but the actual
catalog present in a particular PTA dataset remains unknown. We model this latent catalog as a Poisson point process (PPP) \cite{Kingman1993,DaleyVereJones2003}. A Poisson point process is a
probability model for a set of points scattered through a parameter space, where the number of points is a random variable, and the density of points
is controlled by an intensity function $\mu(s)$. In this construction, the catalog is generated in two steps: first, the number of merger signals \(k\) is drawn from a Poisson distribution with mean
\(\lambda(\Lpop)\); second, conditional on \(k\), the source parameters
\(s_1,\ldots,s_k\) are drawn independently from \(p(s\mid\Lpop)\).

For a catalog \(\cat=\{s_1,\ldots,s_k\}\), the catalog prior is
\begin{equation}
\label{eq:ppp_catalog_prior}
    p(\cat\mid\Lpop)
    =
    e^{-\lambda(\Lpop)}
    \frac{1}{k!}
    \prod_{i=1}^{k}
    \mu(s_i\mid\Lpop).
\end{equation}

When the catalog dimension is truncated at \(K_{\max}\), the count prior is
renormalized on the allowed range,
\begin{equation}
\label{eq:truncated_poisson_prior}
    p(k\mid\lambda(\Lpop),K_{\max})
    =
    \frac{
    {\rm Poisson}\!\left[k\mid\lambda(\Lpop)\right]
    }
    {
    \sum_{j=0}^{K_{\max}}
    {\rm Poisson}\!\left[j\mid\lambda(\Lpop)\right]
    } .
\end{equation}
This truncation is a computational approximation and is safe when
\(P(k>K_{\max}\mid\lambda)\) is negligible over the posterior support.

In the waveform implementation, we use luminosity distance rather than redshift as
the radial source coordinate. The normalized population density transforms as
\begin{equation}
\begin{split}
\label{eq:source_density_DL}
    p(\log_{10}\Mc,D_{\rm L}\mid\Lpop)
    =
    & p(\log_{10}\Mc,z(D_{\rm L})\mid\Lpop)
    \\ &\times
    \left| \frac{\dd z}{\dd D_{\rm L}} \right|.
\end{split}
\end{equation}
The merger time \(t_0\) remains part of \(\vartheta\) and is assigned a
uniform prior over the observing window.

This completes the probabilistic construction of the merger catalog.
We next consider the SGWB predicted by the same source-frame
merger-rate population.

\subsubsection{Stochastic gravitational wave background}
\label{subsubsec:sgwb}

This merger-rate density model also predicts the unresolved SGWB. The SGWB  comes from the accumulated inspiral signal of binaries radiating in the PTA band before merger.

For circular binaries evolving only through gravitational radiation, the
characteristic strain of the SGWB is \cite{Phinney2001,MiddletonDelPozzo2016}
\begin{equation}
\label{eq:hc_middleton}
\begin{split}
    h_c^2(f)
    =
    &\frac{4G^{5/3}}{3\pi^{1/3}c^2}
    f^{-4/3}
    \int \dd\log_{10}\Mc \\
    &\int \dd z\, 
    (1+z)^{-1/3}
    \Mc^{5/3}
    \\
    &\times
    \frac{\dd^3N}
    {\dd V_c\,\dd z\,\dd\log_{10}\Mc}.
\end{split}
\end{equation}

The SGWB calculation includes the broader SMBHB population rather than
being restricted to the explicit merger-catalog domain
\(\mathcal{D}_{\rm m}\). In the present analysis, we evaluate the SGWB
integral over
\[
    10^6\,M_\odot \leq \Mc \leq 10^{11}\,M_\odot,
    \qquad
    0\leq z\leq5.
\]
These are fixed integration bounds rather than sampled population
hyperparameters. They encompass a broader mass and redshift range than the
domain in which individual merger/memory signals are modeled explicitly.

Since \(R(\log_{10}\Mc,z\mid\Lpop)\) is defined per unit source-frame
time, we use
\[
    \left|\frac{\dd t_{\rm R}}{\dd z}\right|
    =
    \frac{1}{(1+z)H(z)}
\]
to convert the source-frame merger rate to the merger density entering the
SGWB calculation. Equation~\eqref{eq:hc_middleton} can therefore be written
directly in terms of the source-frame merger-rate density as
\begin{equation}
\label{eq:hc_from_R}
\begin{split}
    h_c^2 & (f\mid\Lpop)
    =
    \frac{4G^{5/3}}{3\pi^{1/3}c^2}
    f^{-4/3}
    \\
    &\times
    \int \dd\log_{10}\Mc
    \int \dd z\,
    \frac{
        \Mc^{5/3}
        R(\log_{10}\Mc,z\mid\Lpop)
    }{
        (1+z)^{4/3}H(z)
    } .
\end{split}
\end{equation}
Here the factor \((1+z)^{-4/3}\) combines the cosmological time-dilation
factor with the redshift dependence of the emitted inspiral energy
spectrum.

We define the SGWB amplitude at the standard PTA reference frequency
\(f_{\rm yr}=1\,{\rm yr}^{-1}\) as
\begin{equation}
\label{eq:Agwb_from_hc}
    A_{\rm gwb}(\Lpop)
    \equiv
    h_c(f_{\rm yr}\mid\Lpop).
\end{equation}
For the circular, GW-driven population assumed here,
\(h_c(f)\propto f^{-2/3}\), corresponding to
\(\gamma_{\rm gwb}=13/3\). The SGWB is included in the PTA likelihood as an isotropic Hellings--Downs-correlated common red process
\cite{HellingsDowns1983}, with timing-residual power spectral density
\begin{equation}
\label{eq:gwb_residual_psd}
    P(f\mid\Lpop)
    =
    \frac{A_{\rm gwb}^2(\Lpop)}{12\pi^2}
    f_{\rm yr}^{\gamma_{\rm gwb}-3}
    f^{-\gamma_{\rm gwb}},
\end{equation}
where $\gamma_{\rm gwb}=\frac{13}{3}$.Thus, the same population hyperparameters determine both the expected
merger-catalog size \(\lambda(\Lpop)\) and the SGWB amplitude
\(A_{\rm gwb}(\Lpop)\).

\subsection{PTA likelihood implementations}
\label{subsec:pta_likelihood_implementations}

We can use two computational routes to implement the hierarchical likelihood. Both use the same population model, the same merger intensity \(\mu\), and the same PPP catalog prior. They differ only in how the unknown catalog is handled.

\subsubsection{Catalog-marginalized likelihood}
\label{subsubsec:catalog_marginalized_likelihood}

The catalog-marginalized implementation performs separate PTA analyses with
fixed numbers of candidate merger sources. Related fixed-dimension constructions have been used to recover model-marginalized posteriors by performing separate analyses at each allowed model dimension and combining them using their Bayesian evidences
and prior model probabilities~\cite{TongGuttman2025}. Here we apply the same basic evidence-marginalization logic to the unknown merger-catalog size, with an additional population-prior reweighting step described below. This method has two stages.

\paragraph{Step 1: Fixed-\(k\) nested-sampling analyses.}
For each integer \(k=0,1,\ldots,K_{\max}\), we perform a nested-sampling
analysis containing exactly \(k\) deterministic merger signals. Let
\(p_0(s_i)\) denote a reference prior for the population-dependent source
parameters \(s_i\), chosen independently of \(\Lpop\), and let
\(p(\vartheta_i)\) denote the prior on the remaining waveform and geometric
parameters. We use the compact notation
\(s_{1:k}\equiv(s_1,\ldots,s_k)\) and
\(\vartheta_{1:k}\equiv(\vartheta_1,\ldots,\vartheta_k)\) for the parameters
of all \(k\) merger signals. Similarly,
\(\dd s_{1:k}\equiv\prod_{i=1}^{k}\dd s_i\) and
\(\dd\vartheta_{1:k}\equiv\prod_{i=1}^{k}\dd\vartheta_i\).

The reference-prior evidence for the fixed-\(k\) model is
\begin{equation}
\label{eq:fixed_k_ref_evidence}
\begin{split}
    Z_k(d\mid p_0)
    =
    \int&
    p_k(d\mid s_{1:k},\vartheta_{1:k},n)
    \,\pi(n)
    \\
    &\times
    \prod_{i=1}^{k}
    p_0(s_i)\,
    p(\vartheta_i)
    \\
    &\times
    \dd s_{1:k}\,
    \dd\vartheta_{1:k}\,
    \dd n .
\end{split}
\end{equation}
Here \(p_k(d\mid s_{1:k},\vartheta_{1:k},n)\) is the PTA likelihood conditional on exactly \(k\) merger signals, with source and waveform
parameters \(s_{1:k}\) and \(\vartheta_{1:k}\), respectively, and \(n\)
denotes the PTA noise parameters.
The corresponding fixed-\(k\) reference posterior is
\begin{equation}
\begin{split}
    &p_0(s_{1:k},\vartheta_{1:k},n\mid d,k)
    \\
    &\quad =
    \frac{
    p_k(d\mid s_{1:k},\vartheta_{1:k},n)
    \pi(n)
    \prod_{i=1}^{k}p_0(s_i)p(\vartheta_i)
    }{
    Z_k(d\mid p_0)
    }.
\end{split}
\end{equation}
We denote its marginal distribution over the population-dependent source
parameters by
\begin{equation}
    p_0(s_{1:k}\mid d,k)
    =
    \int
    p_0(s_{1:k},\vartheta_{1:k},n\mid d,k)
    \dd\vartheta_{1:k}\dd n 
\end{equation}

\paragraph{Step 2: Population-prior reweighting.}
For population hyperparameters \(\Lpop\), the normalized
source-parameter density \(p(s\mid\Lpop)\) is given by
Eq.~\eqref{eq:normalized_source_density}.

In the merger-only analysis, the PTA likelihood at fixed source and noise
parameters is independent of \(\Lpop\). The evidence under the population
prior can therefore be obtained from the reference analysis using
posterior recycling:
\begin{equation}
\label{eq:fixed_k_reweighted}
\begin{split}
    Z_k(d\mid\Lpop)
    =
    Z_k(d\mid p_0)
    \left\langle
    \prod_{i=1}^{k}
    \frac{
        p(s_i\mid\Lpop)
    }{
        p_0(s_i)
    }
    \right\rangle_{
        p_0(s_{1:k}\mid d,k)
    }.
\end{split}
\end{equation}
The ratio inside the expectation changes the reference source prior into
the population prior. This posterior-recycling step requires sufficient overlap between the reference and population priors. In particular, \(p_0(s)\) must be nonzero wherever \(p(s\mid\Lpop)\) has support relevant to the likelihood. If the population
prior places substantial probability in regions that are poorly represented by the reference posterior, the importance weights
\(p(s\mid\Lpop)/p_0(s)\) can become highly unequal, reducing the effective number of samples contributing to the reweighted likelihood and increasing
its Monte Carlo uncertainty. We therefore choose the reference source prior to cover the source-parameter support of the population models considered in the analysis. Numerically, the expectation is evaluated using the weighted posterior representation returned by the nested sampler. Nested sampling is particularly convenient for this fixed-\(k\) construction because a single analysis provides both the Bayesian evidence \(Z_k(d\mid p_0)\) and the
weighted reference-posterior samples required for the subsequent population-prior reweighting.

The likelihood marginalized over the unknown catalog size is
\begin{equation}
\label{eq:catalog_sum_likelihood}
    p(d\mid\Lpop)
    =
    \sum_{k=0}^{K_{\max}}
    p(k\mid\lambda(\Lpop),K_{\max})
    Z_k(d\mid\Lpop).
\end{equation}
For \(k=0\), no source-prior reweighting is required, so that
\(Z_0(d\mid\Lpop)=Z_0(d\mid p_0)\) in the merger-only model.
The population posterior is therefore
\begin{equation}
\label{eq:catalog_marg_posterior}
    p(\Lpop\mid d)
    \propto
    p(d\mid\Lpop)\,
    \pi(\Lpop).
\end{equation}

The approach becomes less efficient as the allowed catalog size and the dimensionality of the source model increase. A separate nested-sampling
analysis is required for every value of \(k\), and the higher-\(k\) models contain progressively more source parameters. Related transdimensional
studies have noted that such a fixed-dimension strategy can become inefficient because substantial computational effort is spent on model dimensions that are weakly supported by the data~\cite{TongGuttman2025}. Moreover, the posterior-reweighing identity in Eq.~\eqref{eq:fixed_k_reweighted} requires that the PTA likelihood, conditional on the source and noise parameters, be independent of \(\Lpop\). This condition holds for the merger-only implementation. In the joint SGWB+merger model, however, the population-dependent SGWB covariance enters the PTA likelihood directly. The dependence on \(\Lpop\) therefore cannot be represented solely by changing the source prior, and the simple fixed-\(k\) reweighting identity does not apply. For the joint analysis we consequently use the transdimensional implementation.

\subsubsection{Transdimensional likelihood}
\label{subsubsec:explicit_catalog_likelihood}

The second implementation samples the merger catalog explicitly using
reversible-jump Markov-chain Monte Carlo. The sampler moves between models
with different numbers of active merger sources and jointly samples the
population hyperparameters, source parameters, waveform nuisance parameters,
and PTA noise parameters. 

The RJMCMC sampler uses the same truncated count prior defined in
Eq.~\eqref{eq:truncated_poisson_prior}. Conditional on \(k\), each source is drawn from the normalized source prior
\(p(s\mid\Lpop)\) defined in Eq.~\eqref{eq:normalized_source_density}. Thus the sampler implements the same
PPP catalog prior as Eq.~\eqref{eq:ppp_catalog_prior}, restricted to
\(k=0,\ldots,K_{\max}\).

For the joint SGWB+merger analysis, the deterministic merger catalog
enters the mean timing-residual model, while the unresolved SGWB enters the
covariance. The likelihood can be written schematically as
\begin{equation}
\label{eq:rjmcmc_sgwb_merger_likelihood}
\begin{split}
    p(\data\mid s_{1:k},\vartheta_{1:k},\Lpop,n)
    &=
    \mathcal{N}\Big[
    \data - h_{\rm m}(s_{1:k},\vartheta_{1:k}),
    \,
    C_{\rm noise}(n)
    \\
    &\qquad\qquad
    + C_{\rm SGWB}(\Lpop)
    \Big] .
\end{split}
\end{equation}
Here \(h_{\rm m}\) is the summed deterministic merger/memory waveform
contribution, and \(C_{\rm SGWB}(\Lpop)\) is the Hellings--Downs-correlated
covariance implied by the SGWB amplitude derived from the same merger-rate
model.

The full joint posterior sampled by RJMCMC is
\begin{equation}
\label{eq:rjmcmc_sgwb_merger_posterior}
\begin{split}
    &p(\Lpop,n,k,s_{1:k},\vartheta_{1:k}\mid\data)
    \\
    &\quad \propto
    p(\data\mid s_{1:k},\vartheta_{1:k},\Lpop,n)\,
    p(k\mid\lambda(\Lpop),K_{\max})
    \\
    &\qquad \times
    \prod_{i=1}^{k}
    p(s_i\mid\Lpop)\,
    p(\vartheta_i)\,
    \pi(n)\,
    \pi(\Lpop).
\end{split}
\end{equation}

The transdimensional construction is closely related to product-space
sampling, which has been widely used for PTA model selection \cite{NANOGRAV11year,TaylorvanHaasteren2020}. In a standard
product-space analysis, the competing models are embedded in a
fixed-dimensional super-model and an additional model-index parameter
determines which model contributes to the likelihood. Here the discrete
catalog size \(k\) similarly identifies the active source model, but we
sample changes in \(k\) explicitly using reversible-jump birth and death
moves, so that the number of active merger-source parameters changes with
the catalog dimension. Related transdimensional approaches have also been developed using nested sampling. Di Marco et al.~\cite{dimarco2026}, for example, introduce binary indicator variables that switch individual PTA noise processes on and off, while parameters associated with inactive processes are treated as ``ghost'' parameters and excluded from the likelihood evaluation. This allows the nested sampler to explore the different effective model dimensions within a single inference run, providing both parameter posteriors and model probabilities without requiring separate fixed-model analyses.

\section{Simulation and inference}
\label{sec:implementation}

We test the hierarchical framework in two stages. First, we validate the Poisson point-process catalog machinery with a toy merger-rate model in a merger-only simulation. This stage tests whether we can consistently recover both the unknown number of merger signals and the population parameters from the PTA data. Second, we apply the astrophysical merger-rate model to simulations that include a population-dependent SGWB, with and without an individual merger signal.

\subsection{Toy merger-only validation} \label{subsec:toy_setup}

Here we use only a PTA-frame population model rather than the
source-frame astrophysical rate density of Eq.~\eqref{eq:Rsrc_pheno}. The toy catalog intensity is
\begin{equation}
\label{eq:toy_intensity_results}
    \mu_{\rm toy}(s\mid\lambda,\Lambda_{\rm toy})
    =
    \lambda\,p(s\mid\Lambda_{\rm toy})
\end{equation}

where $ \int p(s\mid\Lambda_{\rm toy})\,\dd s=1$, \(\lambda\) is the expected number of merger signals in the PTA dataset,
and
\begin{equation}
    \Lambda_{\rm toy}=(\alpha,\beta,\kappa)
\end{equation}
controls the shape of the source distribution. We write the toy source
coordinates of a merger $i$ as
\begin{equation}
    s_i=(t_{0,i},D_{\mathrm{L},i},\mathcal{M_i},q_i),
\end{equation}
where \(t_0\) is the merger time, \(D_L\) is luminosity distance,
\(\mathcal{M}\) is chirp mass, and \(q\) is the mass-ratio. 

The normalized source distribution factorizes as
\begin{equation}
\begin{split}
\label{eq:toy_source_density}
    p(s\mid\Lambda_{\rm toy})
    &=
    p(t_0)\,
    p(D_L\mid\kappa)\, \\
    &\times p(\mathcal{M}\mid\alpha)\,
    p(q\mid\beta).
\end{split}
\end{equation}
The merger time is uniform over the observing window. The chirp-mass and mass-ratio distributions are normalized power laws,
\begin{equation}
    p(\mathcal{M}\mid\alpha)
    =
    \frac{\mathcal{M}^{-\alpha}}
    {\int_{\mathcal{M}_{\min}}^{\mathcal{M}_{\max}}
    \mathcal{M}^{-\alpha}\,\dd\mathcal{M}}
\end{equation}
and
\begin{equation}
    p(q\mid\beta)
    =
    \frac{q^{-\beta}}
    {\int_{q_{\min}}^{q_{\max}}
    q^{-\beta}\,\dd q}.
\end{equation}
The luminosity-distance distribution is proportional to comoving volume with
an additional redshift-evolution factor,
\begin{equation}
\label{eq:toy_DL_density}
    p(D_\mathrm{L}\mid\kappa)
    =
    \frac{
    \frac{\dd V_c}{\dd D_\mathrm{L}}
    \left[1+z(D_\mathrm{L})\right]^{\kappa}
    }
    {
    \int_{D_{\mathrm{L},\min}}^{D_{\mathrm{L},\max}}
    \frac{\dd V_c}{\dd D_\mathrm{L}'}
    \left[1+z(D_\mathrm{L}')\right]^{\kappa}
    \dd D_\mathrm{L}'
    } .
\end{equation}

The bounds that define the support of the toy source distributions are treated as fixed source hyperparameters. Specifically, \(D_{L,\min}\), \(D_{L,\max}\), \(\mathcal{M}_{\min}\), \(\mathcal{M}_{\max}\), and
\(q_{\min}\), \(q_{\max}\) define the source-parameter domain over which the corresponding distributions are normalized. These quantities are held fixed during the inference. In contrast, the individual quantities \((D_{L,i},\mathcal{M}_i,q_i)\) are source
parameters associated with each merger signal $i$, and are drawn from these distributions. The toy-population priors, fixed source hyperparameters, additional source-parameter priors, and analysis settings are summarized in Table~\ref{tab:toy_priors}.

\begin{table*}
\caption{\label{tab:toy_priors}
Population priors, fixed source hyperparameters, and analysis settings used in the toy merger-only validation. }
\renewcommand{\arraystretch}{1.25}
\begin{ruledtabular}
\begin{tabular}{ccc}
\textbf{Parameter} & \textbf{Prior / Value}  & \textbf{Description} \\ \hline
\multicolumn{3}{c}{\textbf{Toy population hyperparameters}} \\ \hline $\lambda$ & $\mathcal{U}(0.01,3)$ & Expected number of merger signals \\ $\alpha$ & $\mathcal{U}(0,3)$ & Chirp-mass slope \\ $\beta$ & $\mathcal{U}(0,3)$ & Mass-ratio slope \\ $\kappa$ & $\mathcal{U}(-4,4)$ & Redshift-evolution parameter \\ \hline \multicolumn{3}{c}{\textbf{Fixed source hyperparameters}} \\ \hline  $D_{L,\min},D_{L,\max}$ [Mpc] & $100,1000$ & Luminosity-distance bounds \\ $\mathcal{M}_{\min},\mathcal{M}_{\max}$ [$M_\odot$] & $10^8,10^{10}$ & Chirp-mass bounds \\ $q_{\min},q_{\max}$ & $1,7$ & Mass-ratio bounds \\ \hline \multicolumn{3}{c}{\textbf{Additional source-parameter priors}} \\ \hline $t_0$ [MJD] & $\mathcal{U}(53000,58600)$ & Merger Time  \\ $\cos\theta$ & $\mathcal{U}(-1,1)$ & Cosine of Polar angle \\ $\phi$ & $\mathcal{U}(0,2\pi)$ & Azimuthal angle \\ $\psi$ & $\mathcal{U}(0,\pi)$ & Polarization \\ \hline \multicolumn{3}{c}{\textbf{Catalog and noise settings}} \\ \hline Maximum allowed merger signals & $6$ & Catalog-size truncation \\ EFAC & $\mathcal{U}(0.01,10)$ & Error FACtor (white noise) \\
\end{tabular}
\end{ruledtabular}
\end{table*}

For the toy model validation, we simulate a PTA with 25 pulsars distributed uniformly on the sky. The observing span is
\(53000 \leq {\rm MJD} \leq 58600\), corresponding to about 13 years of data. Each pulsar has 1000 times of arrival (TOA), with cadence ranging from a few days to roughly two weeks. The white-noise level is set by
\({\rm EFAC}=1\), and a timing precision of \(100\,{\rm ns}\). The simulated signal is a random catalog of merger events drawn from the
toy PTA-frame population model. 

We analyze the toy simulations using both implementations of the hierarchical likelihood. In the catalog-marginalized implementation, we perform separate PTA analyses containing \(k=0,1,\ldots,6\) merger signals. The \(k=0\) analysis provides the noise-only evidence, while the \(k\geq1\) evidences are computed with \textsc{dynesty} \cite{sergey_koposov_2025_17268284}, using 1000 live points and a stopping criterion \(d\log Z=0.01\). These fixed-\(k\) outputs are then used to construct the population-level
likelihood. For each proposed set of population hyperparameters, the
source posterior samples are importance-reweighted from the reference
source prior to the corresponding population prior using
Eq.~\eqref{eq:fixed_k_reweighted}. The resulting population-dependent
fixed-\(k\) evidences are combined over catalog size through
Eq.~\eqref{eq:catalog_sum_likelihood}. We sample the resulting population
posterior with \textsc{PTMCMCSampler}
~\citep{EllisvanHaasteren2019}. This construction allows the population
likelihood to be evaluated without rerunning the full PTA likelihood at
each point in population-parameter space. Nested sampling is well suited for evidence calculation and multimodal posteriors, but the cost
of sampling generally increases with dimensions. For higher-dimensional source models, a practical extension would be to estimate the fixed-\(k\)
evidences using generalized stepping-stone, which has been proposed for gravitational-wave evidence calculations and recently applied in PTA analyses~\cite{ZahraouiMaturanaRussel2025}.

We also analyze the same toy simulations with the transdimensional implementation using \textsc{Eryn}
\cite{KarnesisKatz2023}. In this case, the sampler explores the
catalog size, the merger-source parameters, the toy population
parameters, and the noise parameters simultaneously. As in the
catalog-marginalized calculation, we allow at most six merger
signals. A closely related transdimensional inference problem arises in searches
for instrumental glitches in LISA data. Muratore et
al.~\cite{MuratoreGair2026} use \textsc{Eryn} to jointly infer an
unknown number of transient glitch components, instrumental noise, and
an astrophysical massive-black-hole-binary signal, combining reversible-jump proposals with parallel tempering and several
complementary in-model proposal mechanisms. Although the physical signals are different, the sampling problem is analogous to the
unknown-merger-catalog problem considered here: efficient exploration requires both continuous parameter updates and transitions between different numbers of transient components.

Motivated by the same general strategy, we use a mixture of proposal types. Group-stretch proposals update the parameters of the currently
active merger signals using nearby states of the sampler ensemble,
while adaptive Gaussian proposals update the source,
population-hyperparameter, and noise parameters using covariance
estimates accumulated during burn-in. Reversible-jump birth and death
proposals add or remove merger signals. Most birth proposals are drawn
from the broad source prior, while a smaller fraction are drawn from an
empirical proposal that is updated during sampling to place new sources
in regions already supported by the likelihood. Parallel tempering is
used to improve exploration of multimodal posterior structure. These
proposals are essential because the complex transdimensional PTA likelihoods can otherwise become trapped in local modes or remain stuck at a fixed source count.

\subsection{Astrophysical SGWB+merger analysis} \label{subsec:astrophysical_setup}

For the astrophysical joint SGWB+merger analysis, we use the phenomenological source-frame merger-rate density of Eq.~\eqref{eq:Rsrc_pheno}, with population hyperparameters \(\Lambda_{\rm pop}\) defined in Eq.~\eqref{eq:population_hyperparameters}. The same \(\Lambda_{\rm pop}\) determines two derived observables: the expected size of the explicitly modeled merger catalog, \(\lambda(\Lambda_{\rm pop})\), and the SGWB amplitude, \(A_{\rm gwb}(\Lambda_{\rm pop})\). As described in Sec.~\ref{subsec:rate_to_pta_observables}, these observables are derived from the same merger-rate density but use different source domains. The SGWB is included in the PTA
likelihood as a Hellings--Downs-correlated common red process with fixed spectral index $\gamma_{\rm gwb}=13/3$. For the explicitly modeled merger/memory catalog, we use $10^8\,M_\odot \leq \Mc \leq 10^{10}\,M_\odot$ and $50\,{\rm Mpc}\leq D_{\rm L}\leq500\,{\rm Mpc}$.
These bounds are
treated as fixed merger-catalog source hyperparameters. Specifically,
\(\mathcal{M}_{\min}\), \(\mathcal{M}_{\max}\),
\(D_{{\rm L},\min}\), and \(D_{{\rm L},\max}\) define the
source-parameter domain \(\mathcal{D}_{\rm m}\) over which the merger-catalog intensity is evaluated and normalized. These quantities are held fixed throughout the inference. In contrast, the individual values \((\mathcal{M}_i,D_{{\rm L},i})\) are source parameters
associated with each merger signal and are inferred from the PTA data. The SGWB integration domain is specified separately in
Sec.~\ref{subsubsec:sgwb}.

We additionally restrict the astrophysical population models used in this analysis to their predicted SGWB amplitude range $-16 \leq \log_{10}A_{\rm gwb}(\Lambda_{\rm pop}) \leq -14.3$.
This range selects population models whose predicted background
amplitudes lie in the regime currently probed by PTA observations. The restriction is applied to the SGWB amplitude derived from \(\Lambda_{\rm pop}\); \(A_{\rm gwb}\) is not sampled as an independent population parameter. A separate restriction is placed on the expected merger-catalog size. We require $\lambda(\Lambda_{\rm pop})\leq2$, which keeps this analysis in a low-count merger regime. This condition acts on the expected catalog size and is not the maximum number of merger signals allowed in an individual transdimensional state. For computational tractability, the joint SGWB+merger recovery allows at most three merger signals, so that the sampled catalog size is \(k=0,1,2,3\). The Poisson count prior is renormalized over these allowed catalog sizes as in Eq.~\eqref{eq:truncated_poisson_prior}.

The joint-analysis simulations use a different PTA setup from the toy validation. We simulate 25 pulsars with $10\,{\rm ns}$ white-noise precision over the observing span $52000 \leq {\rm MJD} \leq 60000$, corresponding to about 22 years of data. We construct two datasets from the same underlying population model. The first contains white noise and an SGWB with amplitude derived from the simulated \(\Lambda_{\rm pop}\). The second contains the corresponding SGWB together with an individual merger drawn from the merger-catalog intensity implied by the same \(\Lambda_{\rm pop}\). This setup
allows us to quantify how much an individual merger signal improves constraints on the merger-rate population beyond the information already provided by the background. The population hyperpriors, fixed mass--redshift support, and analysis settings used in the recovery are summarized in Table~\ref{tab:astrophysical_priors}.

\begin{table*} \caption{\label{tab:astrophysical_priors} Population hyperpriors, fixed merger-catalog source hyperparameters, derived-observable constraints, and fixed analysis settings used for the joint SGWB+merger analysis. The SGWB amplitude $A_{\rm gwb}(\Lambda_{\rm pop})$ and expected merger-catalog size $\lambda(\Lambda_{\rm pop})$ are derived from the population hyperparameters and are not sampled independently. } 
\renewcommand{\arraystretch}{1.25} 
\begin{ruledtabular} 
\begin{tabular}{ccc} \textbf{Parameter} & \textbf{Prior / value} & \textbf{Description} \\ \hline \multicolumn{3}{c}{\textbf{Merger-rate hyperparameters}} \\ \hline $\log_{10}\dot n_0$ [Mpc$^{-3}$ Gyr$^{-1}$ dex$^{-1}$] & $\mathcal{U}(-10,1)$ & Rate normalization \\ $\alpha$ & $\mathcal{U}(-2,2)$ & Chirp-mass slope \\ $\beta$ & $\mathcal{U}(-1,5)$ & Redshift-evolution slope \\ $z_0$ & $\mathcal{U}(0.5,3)$ & Redshift-evolution turnover  \\ $\log_{10}M_*~[M_\odot]$ & $\mathcal{U}(8,9)$ & High-mass turnover  \\ \hline \multicolumn{3}{c}{\textbf{Fixed merger-catalog source hyperparameters}} \\ \hline $\mathcal{M}_{\min},\mathcal{M}_{\max}$ [$M_\odot$] & $10^8,10^{10}$ & Merger-catalog chirp-mass bounds \\ \(D_{{\rm L},\min},D_{{\rm L},\max}\,[{\rm Mpc}]\)
& \(50,500\)
& Merger-catalog luminosity-distance bounds \\ \hline \multicolumn{3}{c}{\textbf{Simulation-selection criteria}} \\ \hline $\log_{10}A_{\rm gwb,\min}, \log_{10}A_{\rm gwb,\max}$ & $-16,-14.3$ & SGWB-amplitude bounds \\ $\lambda_{\max}$ & $2$ & Maximum expected merger-catalog size \\ \hline \multicolumn{3}{c}{\textbf{Catalog and PTA settings}} \\ \hline Maximum allowed merger signals & $3$ & Catalog-size truncation \\ $\gamma_{\rm gwb}$ & $13/3$ & SGWB spectral index \\ EFAC & $1$ & Error FACtor (white noise) \\ \end{tabular} \end{ruledtabular} \end{table*}

\section{Results}

\subsection{Toy merger-only population}
\label{sec:results_toy}

\begin{figure*}[t]
    \centering
    \begin{subfigure}{0.32\textwidth}
    \includegraphics[width=\textwidth]{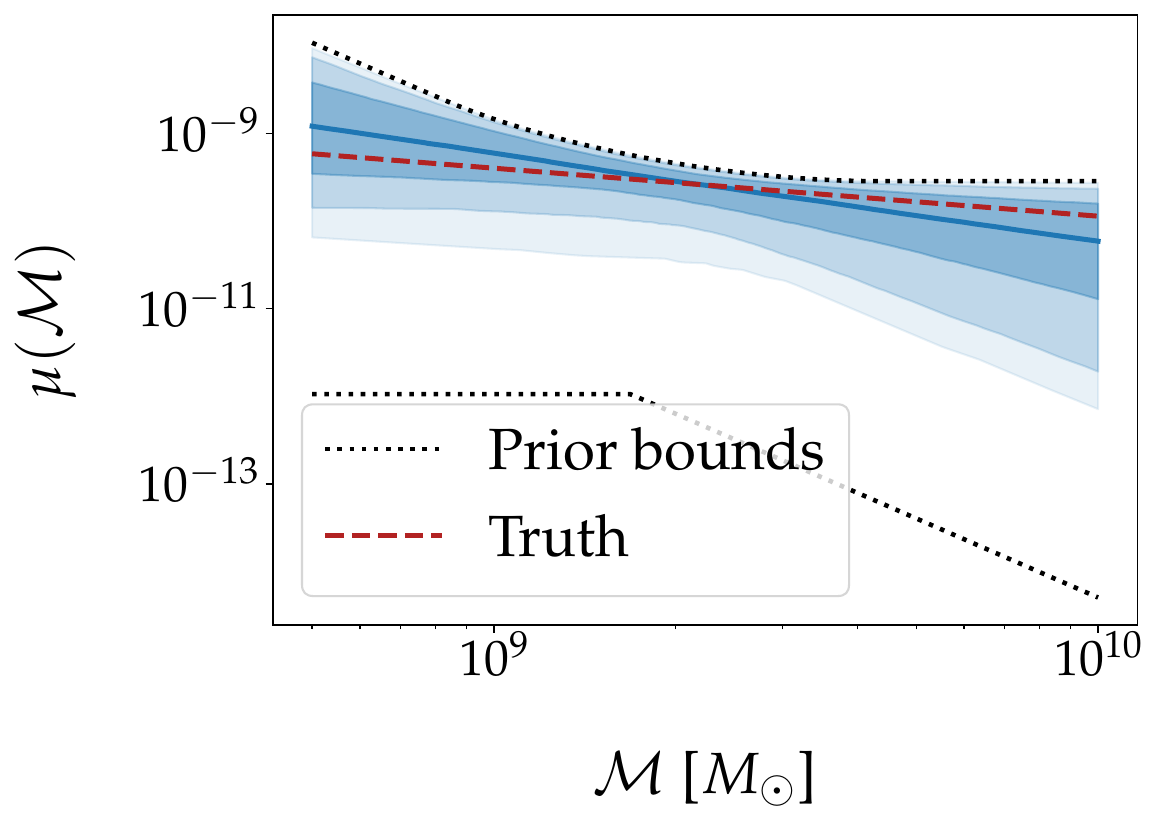}
    \caption{}
    \label{fig:merger_mc}
  \end{subfigure}
  \hfill
  \begin{subfigure}{0.32\textwidth}
    \includegraphics[width=\textwidth]{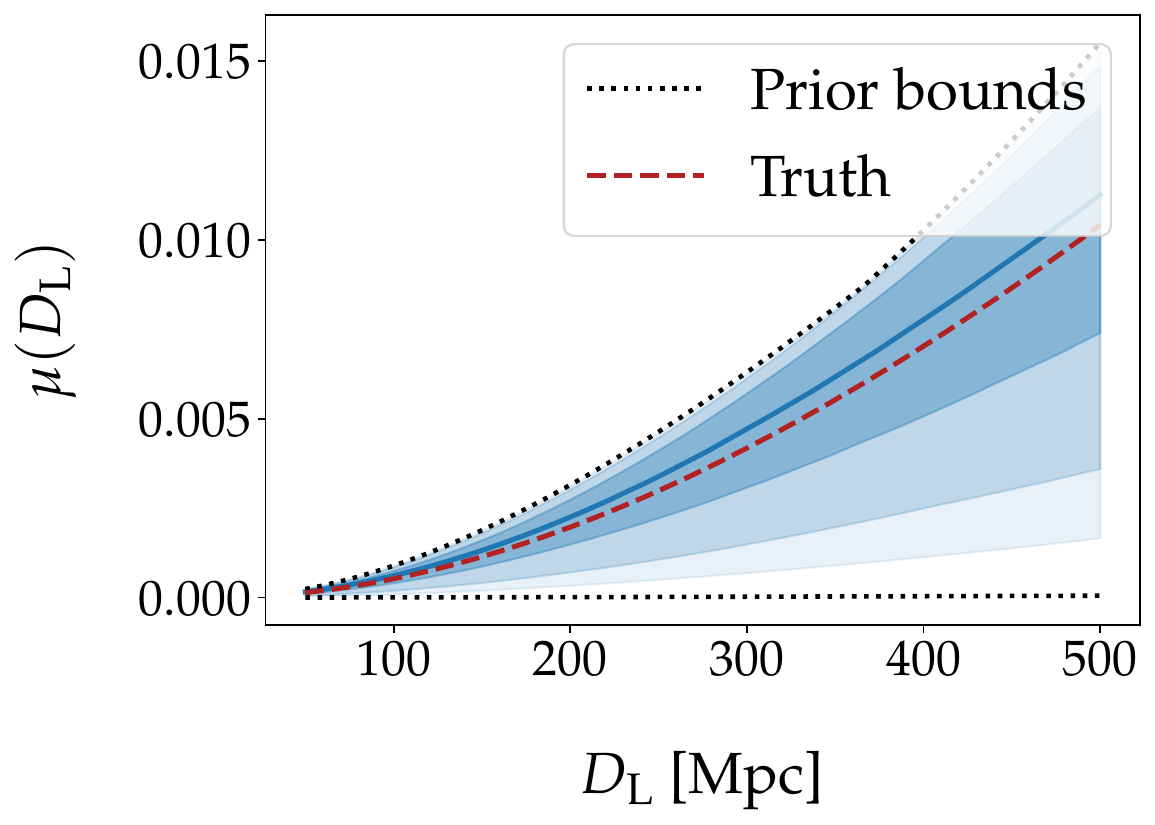}
    \caption{}
    \label{fig:merger_dl}
  \end{subfigure}
  \begin{subfigure}{0.32\textwidth}
    \includegraphics[width=\textwidth]{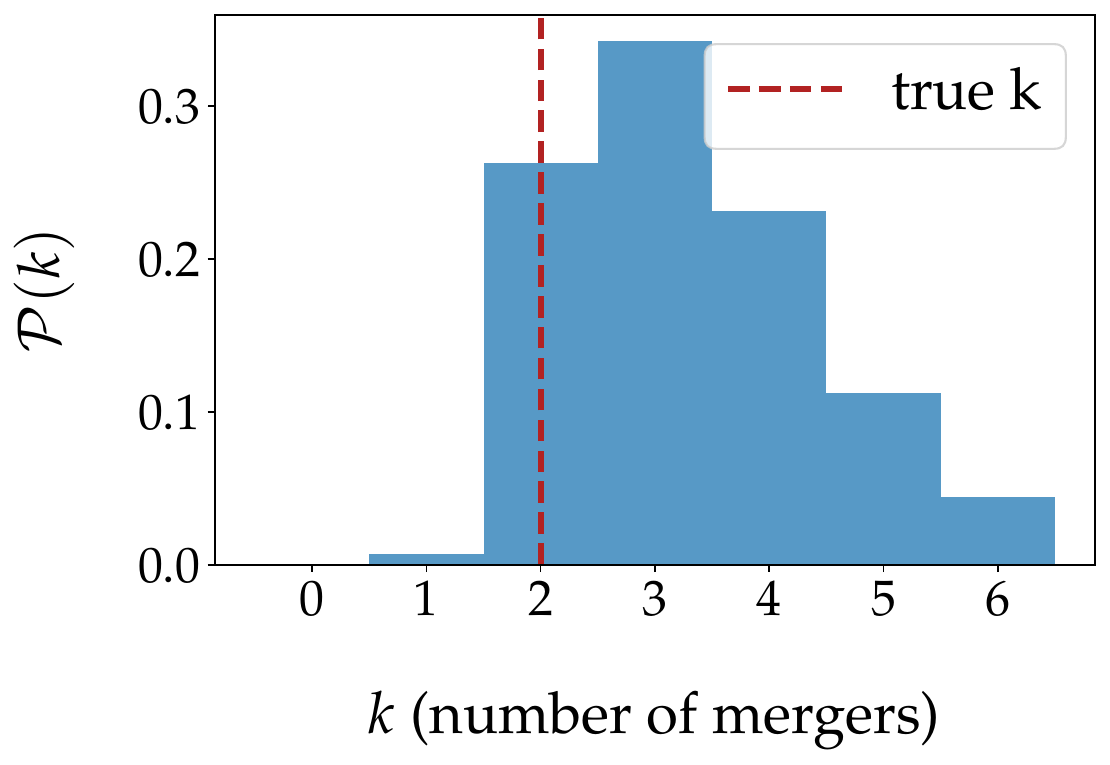}
    \caption{}
    \label{fig:k_posterior}
  \end{subfigure}
    \caption{\justifying
    Toy merger-only validation of the Poisson point-process population model.
    Panels (a) and (b) show the recovered marginal PTA-frame merger-rate
    distributions (catalog intensities) in chirp mass and luminosity distance,
    respectively. The blue solid curve denotes the posterior median, the blue
    shaded regions show the 68\%, 95\%, and 99.7\% posterior credible bands, the
    black dotted curves mark the lower and upper envelopes implied by the
    hyperprior bounds, and the red dashed curve shows the true simulated population.
    Panel (c) shows the posterior probability mass $\mathcal{P}(k)$ for the catalog size $k$, with the dashed vertical line marking the simulated number of mergers in the data. 
    }
    \label{fig:toy_validation}
\end{figure*}

We first present the results of the toy merger-only validation described
in Sec.~\ref{subsec:toy_setup}. Figure~\ref{fig:toy_validation} shows a recovery using the transdimensional likelihood for one simulation. The simulated population for this realization has \(\lambda=1.86\), \(\alpha=0.55\), \(\beta=1.49\), and \(\kappa=0.96\), and the simulated catalog contains \(k_{\rm true}=2\) merger signals. The simulated source parameters for this realization are given in Appendix~\ref{app:toy_validation_details}. Figures~\ref{fig:merger_mc} and \ref{fig:merger_dl} show the inferred marginal catalog intensity distributions as functions of chirp mass and luminosity distance. The red dashed curves show
the underlying true PTA-frame merger rate population, while the blue curves and shaded regions show the posterior median and credible bands. The black dotted curves mark the lower and upper envelopes implied by the
population priors and fixed source hyperparameters. These panels test whether the population shape is recovered after marginalizing over the unknown catalog. Fig.~\ref{fig:k_posterior} shows the posterior distribution for the number of mergers in the dataset. The dashed vertical line marks the simulated source count. 

The same toy setup was also validated with the catalog-marginalized
fixed-\(k\) implementation. In that case, the fixed-source-number evidences
computed with nested sampling were combined through the finite catalog sum, rather than sampling the catalog dimension directly. The two implementations give consistent recovery of the toy population parameters, providing a useful
cross-check that the catalog marginalization and the transdimensional sampler are implementing the same hierarchical model.

Additional diagnostics for this representative toy realization,
including the population-hyperparameter posterior, the source-parameter
recovery, and a convergence diagnostic for the population parameters and
catalog size, are shown in Appendix~\ref{app:toy_validation_details}.

We next test whether the population inference is statistically consistent over an
ensemble of simulations using probability--probability (PP) plots. For each
simulation and each population hyperparameter, we evaluate the posterior
cumulative distribution function (CDF) at the simulated parameter value. This
gives the posterior quantile of the simulated value. We then construct the
empirical cumulative distribution function (ECDF) of these posterior quantiles
across the ensemble of simulations. If the inference is calibrated, the
posterior quantiles are uniformly distributed between 0 and 1, so the ECDF
follows the diagonal. Systematic deviations from the diagonal indicate
undercoverage or overcoverage.

\begin{figure*}[t]
    \centering
    \includegraphics[width=0.95\linewidth]{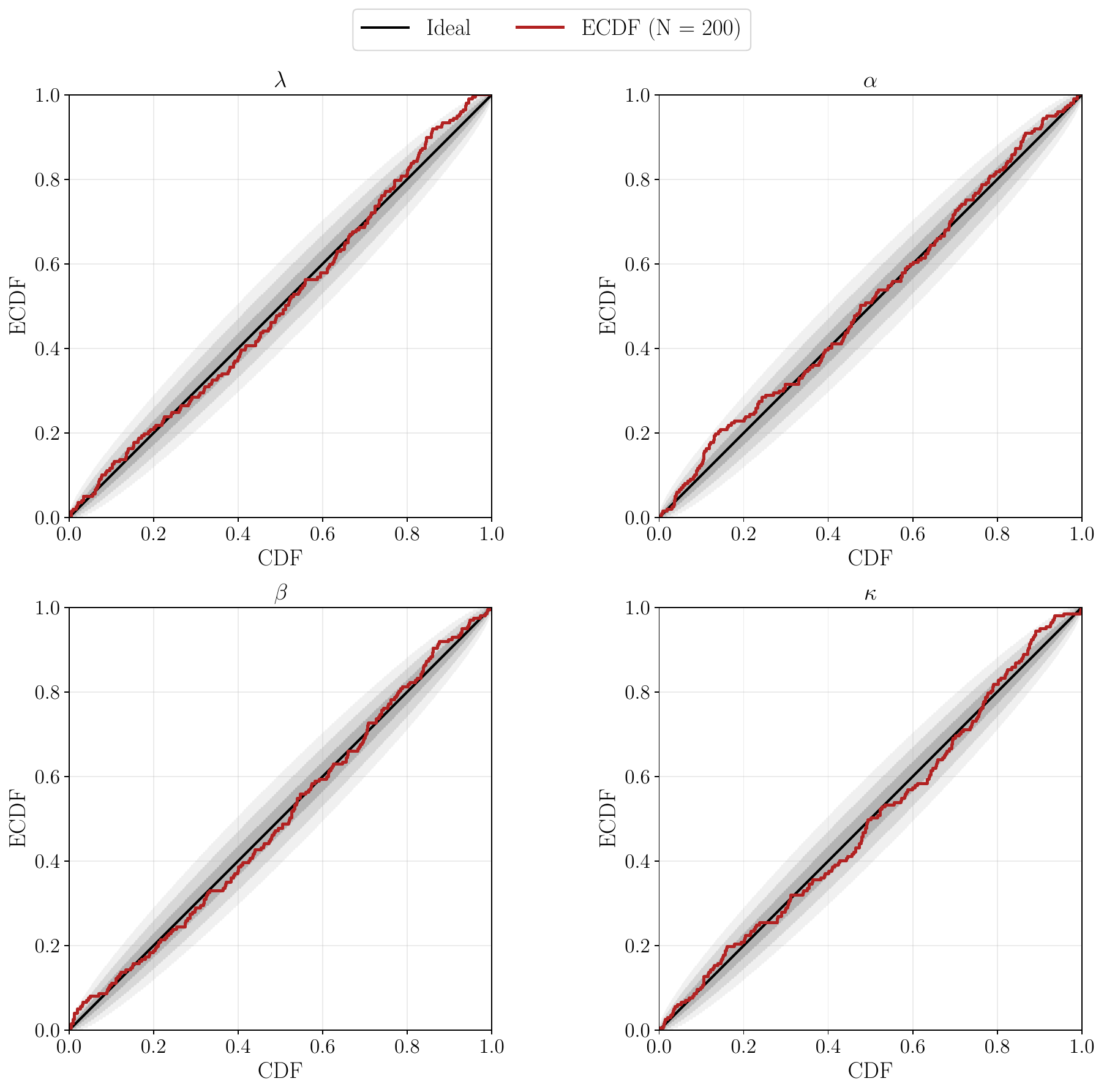}
    \caption{\justifying
    PP-plot calibration test for the toy merger-only population inference using
    200 simulations. For each population hyperparameter, the horizontal axis
    shows the posterior CDF evaluated at the simulated value, and the vertical
    axis shows the ECDF of these values across the simulation ensemble. The black
    diagonal denotes the expectation for calibrated inference, the red curve is
    the measured ECDF, and the gray regions show the expected 1-, 2-, and
    3-\(\sigma\) finite-simulation bands.
    }
    \label{fig:toy_pp}
\end{figure*}

Figure~\ref{fig:toy_pp} shows the resulting calibration test for 200
simulations. Each panel corresponds to one population hyperparameter. The
horizontal axis shows the posterior CDF value of the simulated parameter,
while the vertical axis shows the ECDF of those values over the simulation
ensemble. The red curve is the measured ECDF, the black diagonal is the
expectation for calibrated inference, and the gray bands show the expected
finite-sample fluctuations.

\subsection{Merger-rate inference from the joint SGWB and merger analysis}
\label{sec:results_joint}

After validating the toy population model, we apply the framework to the phenomenological source-frame merger-rate model of Eq.~\eqref{eq:Rsrc_pheno}. We compare the SGWB-only and SGWB+merger datasets described in Sec.~\ref{subsec:astrophysical_setup}. Both datasets are generated from the same underlying merger-rate population, so that their comparison isolates the additional population information carried by an individual merger signal.

As discussed in Sec.~\ref{subsec:astrophysical_setup}, we restrict the
astrophysical population models to a range of SGWB amplitudes relevant to
current PTA observations. Because the same population hyperparameters
determine both the SGWB amplitude and the merger rate, populations in this
regime generally predict a small number of individual mergers within the
finite PTA observing span and the explicitly modeled merger-catalog domain.
Individual merger events are therefore rare in this PTA-frame catalog. To
study the population information available when such an event is present,
we condition the realization used here to contain exactly one merger signal. For the selected realization, the injected population predicts an expected
merger-catalog size of \(\lambda(\Lpop)=0.0193\). The simulated background has \(\log_{10}A_{\rm gwb}=-14.352\), corresponding to
\(A_{\rm gwb}\simeq4.45\times10^{-15}\), with fixed \(\gamma_{\rm gwb}=13/3\). For this value of \(\lambda\), an exactly
one-merger catalog is a rare outcome, with probability approximately
\(1.9\%\) under the truncated count model used here. The selected
realization is therefore deliberately atypical rather than a representative
unconditional draw. It provides a controlled test of the additional
population information obtained when an individual merger is present.

\begin{figure}[t] 
\centering \includegraphics[width=0.98\linewidth]{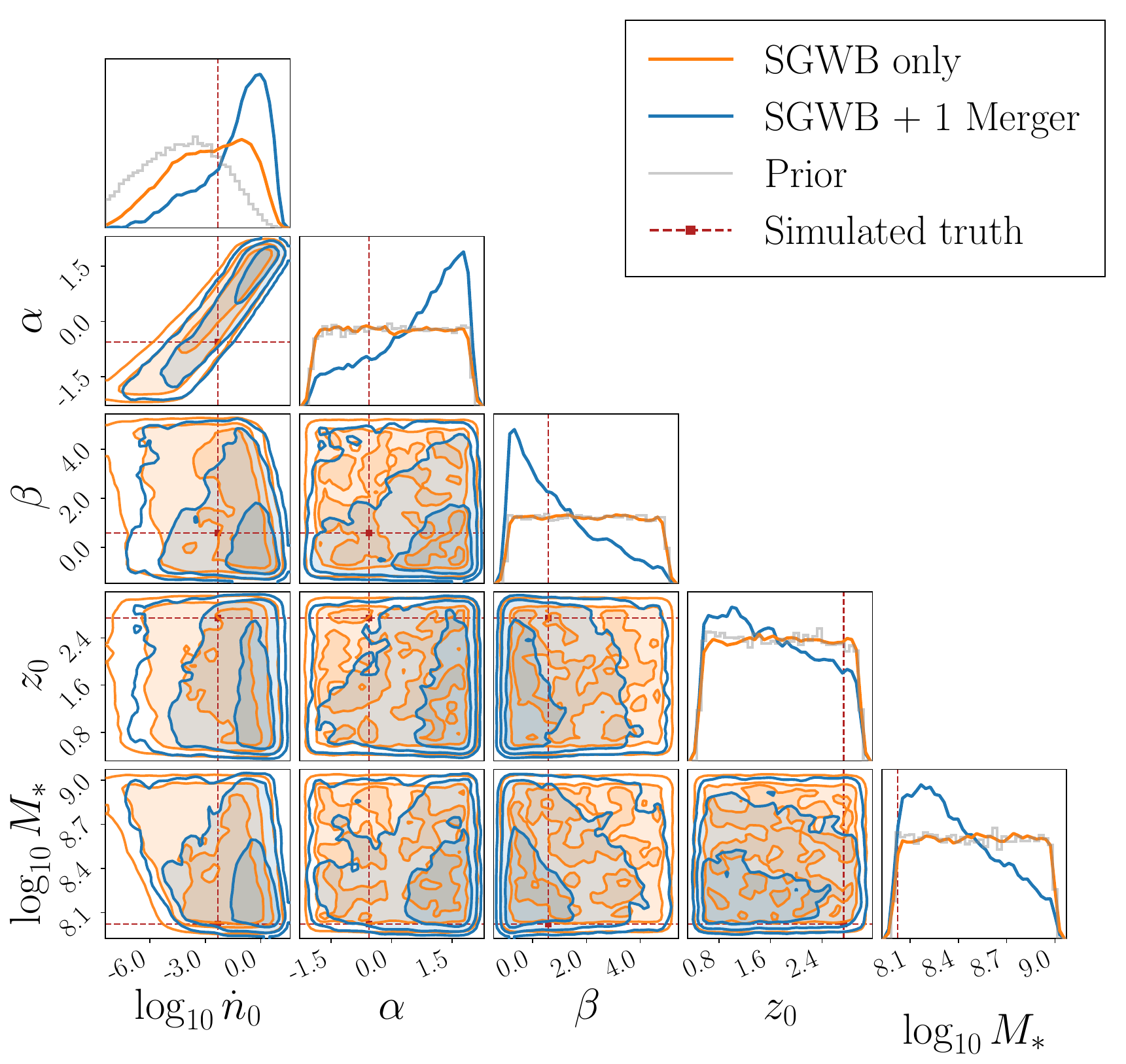} 
\caption{\justifying Population-hyperparameter posteriors for the SGWB-only and SGWB+merger analyses of the representative realization conditioned to contain one merger signal. Orange contours show the SGWB-only recovery, blue contours show the SGWB+merger recovery, and red dashed lines mark the simulated merger-rate hyperparameters. Both datasets are generated from the same underlying population, so the comparison isolates the additional population information contributed by the merger signal. } 
\label{fig:joint_corner} 
\end{figure}

Figure~\ref{fig:joint_corner} shows the resulting population-hyperparameter posteriors. The orange contours show the SGWB-only recovery, while the blue contours show the recovery from the SGWB+merger dataset generated from the same underlying population. The red dashed lines mark the simulated merger-rate hyperparameters. The SGWB-only posterior remains broad because different combinations of the merger-rate normalization, mass dependence, and redshift evolution can produce similar values of \(A_{\rm gwb}\). Conditioning on the presence of one merger and including its measured source
properties substantially changes several of these degeneracies. For this
realization, the one-merger analysis favors larger values of the rate
normalization and \(\alpha\). More generally, the additional information
provided by the merger reduces the range of population models consistent
with the data and leads to tighter constraints on several of the
population-level hyperparameters. The SGWB and individual merger signal consequently constrain different projections of the same merger-rate population: the SGWB constrains the population-integrated unresolved inspiral signal, whereas the merger signal directly probes the rare, nearby, high-mass part of the population. 

\begin{figure*}[t] \centering  \includegraphics[width=0.98\linewidth]{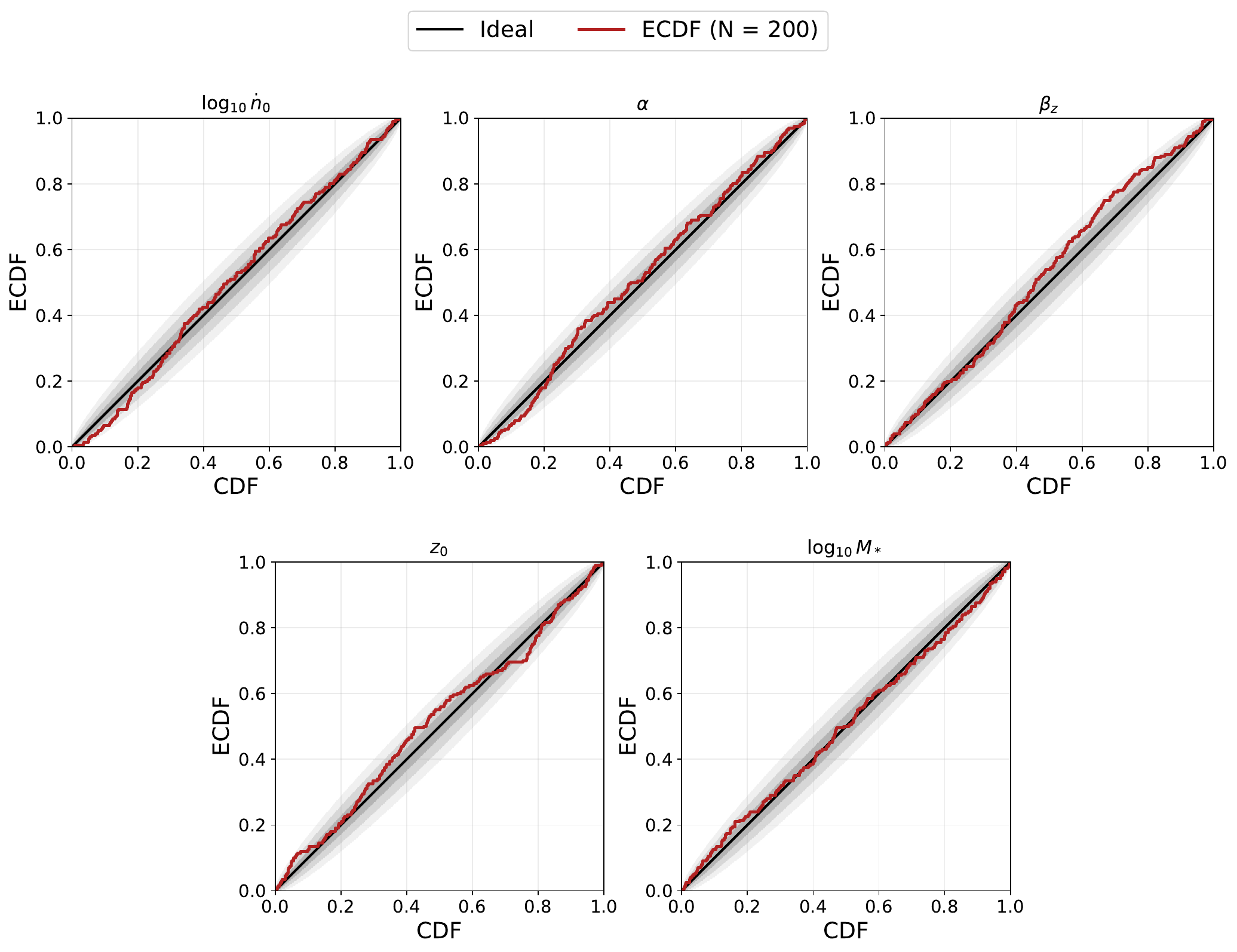} \caption{\justifying PP-plot calibration test for the five merger-rate hyperparameters in the conditioned one-merger analysis. Each simulated realization contains exactly one merger signal and is analyzed with the corresponding one-merger model. The black diagonal denotes ideal statistical calibration, the red curve is the measured ECDF, and the gray regions show the $1-,2-,$ and $3-\sigma$ credible regions. Curves consistent with the diagonal indicate correct posterior coverage for the merger-rate hyperparameters. } \label{fig:joint_pp} 
\end{figure*}

We next test whether the astrophysical population inference is
statistically consistent when an individual merger is present. For this calibration test, we use an ensemble of simulations containing exactly one merger signal and analyze each realization with the corresponding one-merger
model. Figure~\ref{fig:joint_pp} shows the PP plots for the five merger-rate
hyperparameters
\(\log_{10}\dot n_0\), \(\alpha\), \(\beta\), \(z_0\), and
\(\log_{10}M_*\). The empirical curves remain consistent with the
diagonal within the expected finite-simulation fluctuations, indicating
statistically consistent recovery of the population hyperparameters in
the conditioned one-merger analysis.

\section{Discussion}
\label{sec:discussion}

The main purpose of this work is to show that PTA merger-rate inference can be formulated directly as a hierarchical population problem. In this framework, the
SGWB, individual mergers, weak candidates, and nondetections are not separate statistical products to be interpreted afterward. They are different pieces of information about the same underlying SMBHB merger-rate population. The Poisson point-process construction provides a natural way to
include the unknown source catalog in the likelihood, so that the number of merger signals is inferred from the data rather than fixed by a detection threshold.

The two likelihood implementations have different computational
advantages. The catalog-marginalized method provides a useful validation
route because the fixed-source-count analyses can be performed
independently and subsequently combined with the population prior.
However, its computational cost grows as more catalog sizes and
higher-dimensional source models must be analyzed, and the
population-prior reweighting is applicable only when the fixed-source PTA
likelihood is independent of the population hyperparameters. This
condition fails in the joint SGWB+merger problem because the SGWB
covariance depends directly on \(\Lpop\).

For this reason, the transdimensional likelihood is the more direct route for
the joint SGWB-plus-merger problem. It samples the population hyperparameters,
the source count, the active merger catalog, and the noise parameters
simultaneously, while the SGWB amplitude is recomputed from the same
population model. This approach is more flexible, but it introduces the usual
challenges of transdimensional inference. Because the merger signal can be weak, several low-amplitude source
components can sometimes mimic one louder merger, or one merger signal can
be partially represented by multiple sources. Without careful proposal design and sufficient prior penalty for
unnecessary catalog complexity, the sampler can become trapped in local modes
or spend too much time in incorrect source dimensions. This is analogous to
the source-confusion and model-dimension challenges that appear in LISA
global-fit problems. We therefore do not rely on a single ensemble proposal to explore this posterior. Within a fixed catalog dimension, we use a mixture of proposal mechanisms: the group-stretch move updates individual merger-source blocks,
while adaptive Gaussian proposals provide complementary local and multivariate updates for the source and population parameters. Changes in catalog dimension are explored separately through reversible-jump birth--death proposals, using both broad prior-based births and an empirical
birth distribution adapted during sampling. Parallel tempering provides an
additional mechanism for moving between separated posterior modes. This
mixture of complementary proposals is important for robust exploration of
the transdimensional posterior. In particular, the group-stretch proposal is applied blockwise to the six-dimensional parameter vector of an individual merger signal rather than to the full joint parameter space. For the present proof-of-concept joint analysis, we restrict the
transdimensional catalog to at most three merger signals. This
computational truncation is adequate for the low-count realization studied
here, but analyses of populations supporting larger merger counts will
require a correspondingly larger allowed catalog.

The present implementation is intentionally a proof of concept. For the
individual merger channel, we use a simplified burst-with-memory model in
which the memory appears as a step in strain and a ramp in timing residuals.
For future work the complete accurate signal model needs to be used from Ref.~\cite{TomsonGoncharovHaasteren2026}.

The SGWB treatment is also simplified. In the current implementation, the
source-frame merger-rate model is mapped to a single SGWB amplitude
\(\Agwb(\Lpop)\) with fixed spectral index. This is sufficient
for testing whether the same population hyperparameters can control both the
SGWB amplitude and the expected merger catalog, but it does not use all of the
spectral information available in PTA data. Future versions of the framework
could instead use a free-spectrum or piecewise spectral representation, so
that the full frequency dependence of the SGWB contributes to the population
likelihood. This will become increasingly important as PTA measurements improve,
because the detailed spectral shape can help distinguish SMBHB astrophysics
from other possible origins of the nanohertz background
\cite{NG_15_HOLODECK}.

The astrophysical population model can also be made more informative. The
phenomenological merger-rate density used here is useful for validating the
hierarchical machinery, but it compresses the astrophysics into a small number
of mass and redshift parameters. More physical models connect the SGWB and
merger-rate population to galaxy stellar mass functions, galaxy pair fractions,
merger timescales, black-hole--host-galaxy relations, binary hardening,
environmental coupling, and eccentricity \cite{ChenSesana2019}. Such
models can be inserted into the same framework by replacing the simple
\(R(\Lpop)\) used here with a more physical merger-rate
population.

This framework also points toward a unified treatment of the SGWB, continuous
waves, and merger events. PTAs have now entered an era in which multiple datasets show evidence for the SGWB, and targeted searches for individually
resolvable SMBHB continuous waves are becoming increasingly important. If
bright continuous-wave sources begin to appear together with the unresolved
background, the separation between ``resolved'' and ``unresolved'' sources
will itself become part of the population-inference problem. One possible
extension is to adapt the resolved/unresolved population-splitting logic used
in recent global-fit work: the unresolved background is treated as the
collective contribution of subthreshold sources, while the brightest sources
are modeled explicitly as part of the same population \cite{ToubianaGair2026,CriswellBanagiri2026}.
In a PTA application, this would amount to modeling the SGWB as the unresolved
part of a Poissonian SMBHB population, while introducing a population-dependent
resolvability boundary in continuous-wave parameter space. Machine-learning emulators may provide a practical way to make this extension
computationally feasible. Recent LISA work has shown how simulated source
catalogs and reconstructed foreground spectra can be used to train a neural
posterior estimator for population hyperparameters \cite{ToubianaGair2026}.
In the PTA context, \citet{LaalTaylor2025} demonstrated that a normalizing-flow emulator can learn the population-dependent distribution of SGWB spectra from simulated
SMBHB populations. Building on this approach, an extension would be to emulate the joint mapping from SMBHB population parameters to both the unresolved SGWB spectrum and the distribution of resolvable continuous-wave sources.

A particularly useful alternative has recently been developed by
\citet{GoncharovSato-Polito2026}. Rather than sampling an
unknown catalog of continuous-wave sources, their construction treats the
brightest SMBHB as the source most likely to become individually resolvable
and marginalizes the remaining binaries into the unresolved background.
Crucially, the continuous-wave source and the SGWB are not assigned
independent phenomenological priors: the probability distribution of the
brightest source and that of the remaining background are derived from the
same underlying Poisson SMBHB population. The method therefore retains the
population coupling between resolved and unresolved sources without requiring
a transdimensional search over an arbitrary number of continuous-wave
signals. This provides a particularly useful route for extending the framework
developed here. The brightest source carries much of the information
associated with the high-strain tail of the SMBHB population and is also the
source most likely to become individually resolvable. Modeling this source
explicitly, while statistically marginalizing the much larger population of
weaker binaries into the SGWB, therefore captures the leading additional
population information available from continuous waves. At the same time,
the calculation can be implemented within the standard fixed-dimensional PTA
likelihood, with the connection between the SGWB and the continuous-wave
source entering through their common hierarchical population prior
\cite{GoncharovSato-Polito2026}. This is considerably simpler than introducing
a reversible-jump sampler for an unknown number of resolvable binaries.
The more general resolved/unresolved population-splitting and emulator-based
approaches discussed above remain valuable if future PTA datasets contain
several comparably informative resolvable sources or require substantially
more complicated source populations.

\section{Conclusions}
\label{sec:conclusion}

We have developed a hierarchical Bayesian framework for inferring the SMBHB
merger-rate population directly from PTA data. Instead of treating
deterministic-search upper limits and SGWB measurements as separate products to
be interpreted after the fact, the framework places the source-frame
merger-rate model inside the PTA likelihood. The unknown catalog of individual
SMBHB merger signals is modeled as a Poisson point process, so that the number
of merger signals is inferred from the data rather than fixed by a detection
threshold.

We presented two computational implementations of this hierarchical idea. The
first marginalizes over the unknown catalog by combining fixed-source-count
evidences with the Poisson catalog prior. This catalog-marginalized route is
useful for merger-only validation problems, where the population
hyperparameters change the source prior but not the fixed-source PTA
likelihood. The second implementation samples the catalog explicitly with
RJMCMC, jointly exploring the population hyperparameters, source count, source
parameters, and PTA noise parameters. This transdimensional route is the more
direct implementation for the joint SGWB-plus-merger problem, because the SGWB
covariance depends on the same population hyperparameters.

We validated the Poisson point-process machinery using a toy merger-only
population. The toy study shows that the framework can recover both the
marginal source distributions and the unknown merger count. Probability--
probability tests over an ensemble of simulations show that the inferred
population hyperparameters have the expected coverage. This validates the
basic unknown-catalog inference before applying the method to an astrophysical
merger-rate model.

We then applied the framework to a phenomenological astrophysical merger-rate density in which the same population hyperparameters determine both the SGWB amplitude and the expected merger catalog. In controlled simulations, adding an individual merger signal to the SGWB analysis changes the population posterior and can improve constraints
on the merger-rate hyperparameters. This demonstrates the central point of the analysis: the SGWB and individual merger/memory signals carry complementary information about the same underlying SMBHB population.

The present implementation is a proof of concept. The merger signal is modeled with a simplified burst-with-memory waveform, and the SGWB is compressed to a power-law amplitude with fixed spectral index. Future work will replace the burst approximation with a complete SMBHB merger waveform, use more of the frequency-dependent SGWB information, and extend the same hierarchical population model to include resolvable continuous-wave sources. This would lead to a unified PTA population framework in which the SGWB, individual continuous waves, merger/memory events, and nondetections all constrain the same astrophysical SMBHB merger-rate population.

\acknowledgements

We thank Martina Muratore for insightful scientific discussions and for
guidance on the \textsc{Eryn} transdimensional sampler, including
proposal design, sampler settings, and convergence diagnostics. This work is supported by the Max Planck Gesellschaft (MPG) and the ATLAS cluster computing team at AEI Hannover.

We made use of the following code.
For simulating pulsar timing datasets with realistic observation spans, noise properties, and cadence, we use the \textsc{pta\_replicator} at \href{https://github.com/bencebecsy/pta_replicator.git}{github.com/bencebecsy/pta\_replicator} with an extension to include our SMBHB merger with memory parameters. 

The Bayesian analysis is performed using the enterprise framework~\citep{EllisVallisneri2020}, the core data analysis software for pulsar timing arrays for computing likelihoods and posteriors. 
For the catalog-marginalized implementation, we use
\textsc{dynesty} \href{https://github.com/joshspeagle/dynesty.git}{github.com/joshspeagle/dynesty.git} to compute the fixed-\(k\) nested-sampling evidences, and \textsc{PTMCMCSampler} \href{https://github.com/nanograv/PTMCMCSampler.git}{github.com/nanograv/PTMCMCSampler.git} for the population-level reweighting and hyperparameter sampling. For the transdimensional implementation, we use
\textsc{Eryn} \href{https://github.com/lisa-analysis-tools/Eryn.git}{github.com/lisa-analysis-tools/Eryn.git}, which provides ensemble and reversible-jump MCMC tools for sampling across different catalog dimensions.

\appendix
\label{app:appendix}

\section{Toy-population validation details}
\label{app:toy_validation_details}

This appendix provides additional diagnostics for the representative toy
merger-only realization discussed in Sec.~\ref{sec:results_toy}. The toy
population is the PTA-frame model defined in
Eq.~\eqref{eq:toy_intensity_results}. The simulated population hyperparameters for this realization are
\[
    \lambda=1.85847, 
    \alpha=0.54793,
    \beta=1.49214,
    \kappa=0.96077 .
\]
The simulated catalog contains two merger signals. Their simulated parameters
are listed in Table~\ref{tab:toy_injected_catalog}. Additional figures below show the recovery of the population hyperparameters and source parameters, together with convergence diagnostics for the transdimensional sampler.

\begin{table*}
\caption{\label{tab:toy_injected_catalog}
Simulated merger signals for the representative toy validation realization
shown in Fig.~\ref{fig:toy_validation}. The parameters are the simulated values used to generate the PTA dataset.}
\renewcommand{\arraystretch}{1.25}
\begin{ruledtabular}
\begin{tabular}{ccccccccc}
\textbf{Event} &
\(t_0\) [MJD] &
\(D_L\) [Mpc] &
\(\mathcal{M}\,[M_\odot]\) &
\(q\) &
\(\theta\) &
\(\phi\) &
\(\psi\) &
\(h_{\rm mem}\) \\ \hline
1 &
54388.419 &
427.761 &
\(5.965\times10^9\) &
5.386 &
1.723 &
6.018 &
2.506 &
\(1.015\times10^{-14}\) \\
2 &
57820.904 &
341.978 &
\(2.234\times10^9\) &
1.251 &
1.196 &
5.455 &
2.984 &
\(6.010\times10^{-15}\) \\
\end{tabular}
\end{ruledtabular}
\end{table*}

\begin{figure}[t]
    \centering
    \includegraphics[width=0.95\linewidth]{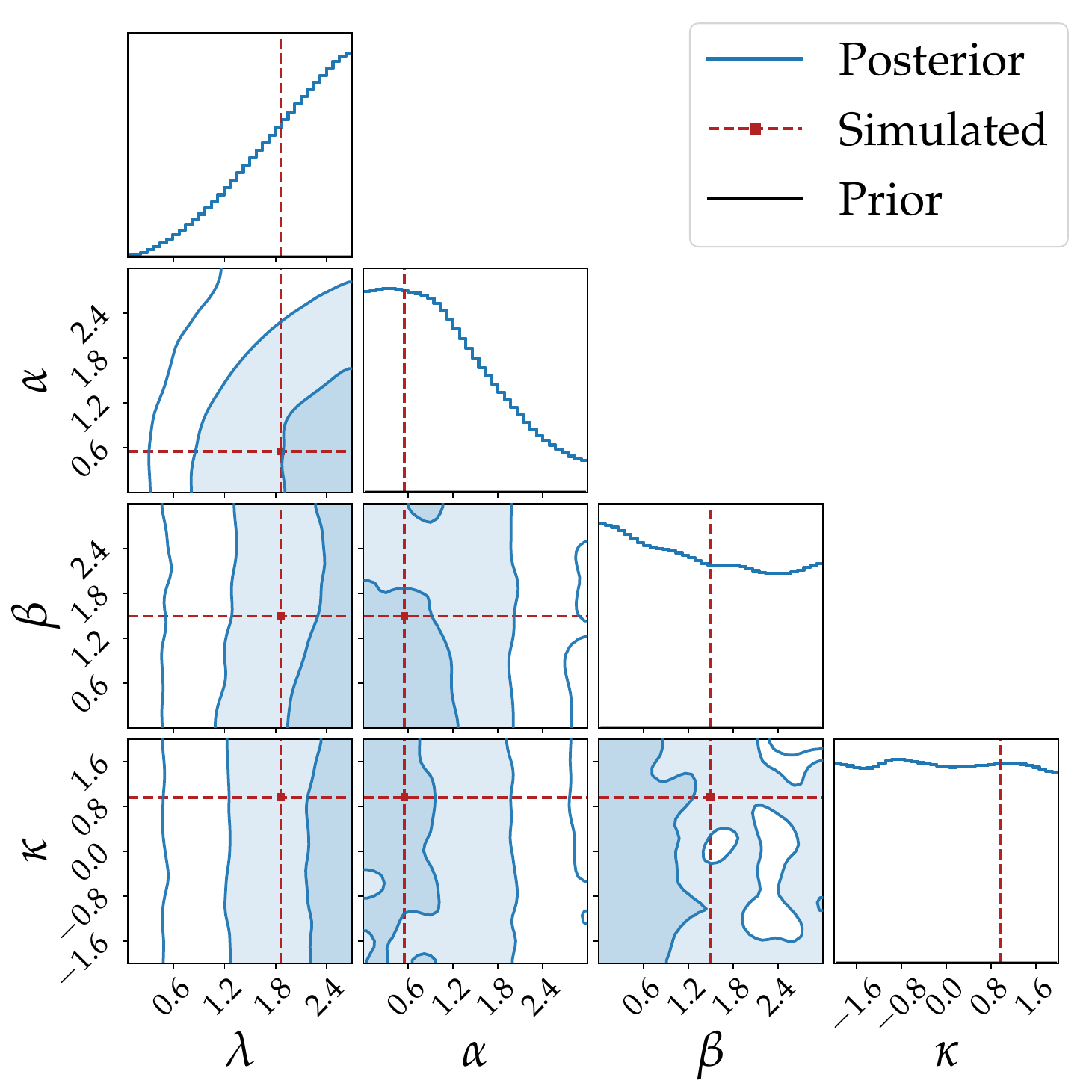}
    \caption{\justifying
    Posterior on the toy population hyperparameters
    \((\lambda,\alpha,\beta,\kappa)\) for the representative validation
    realization. The blue contours and one-dimensional marginals show the recovered
    posterior, while the red dashed lines mark the simulated values. This figure
    shows the full joint recovery of the toy population model for a single
    realization.
    }
    \label{fig:toy_hyper_corner_app}
\end{figure}

Figure~\ref{fig:toy_hyper_corner_app} shows the posterior on the four toy population hyperparameters for this realization. 

\begin{figure*}[t]
    \centering
    \includegraphics[width=0.92\linewidth]{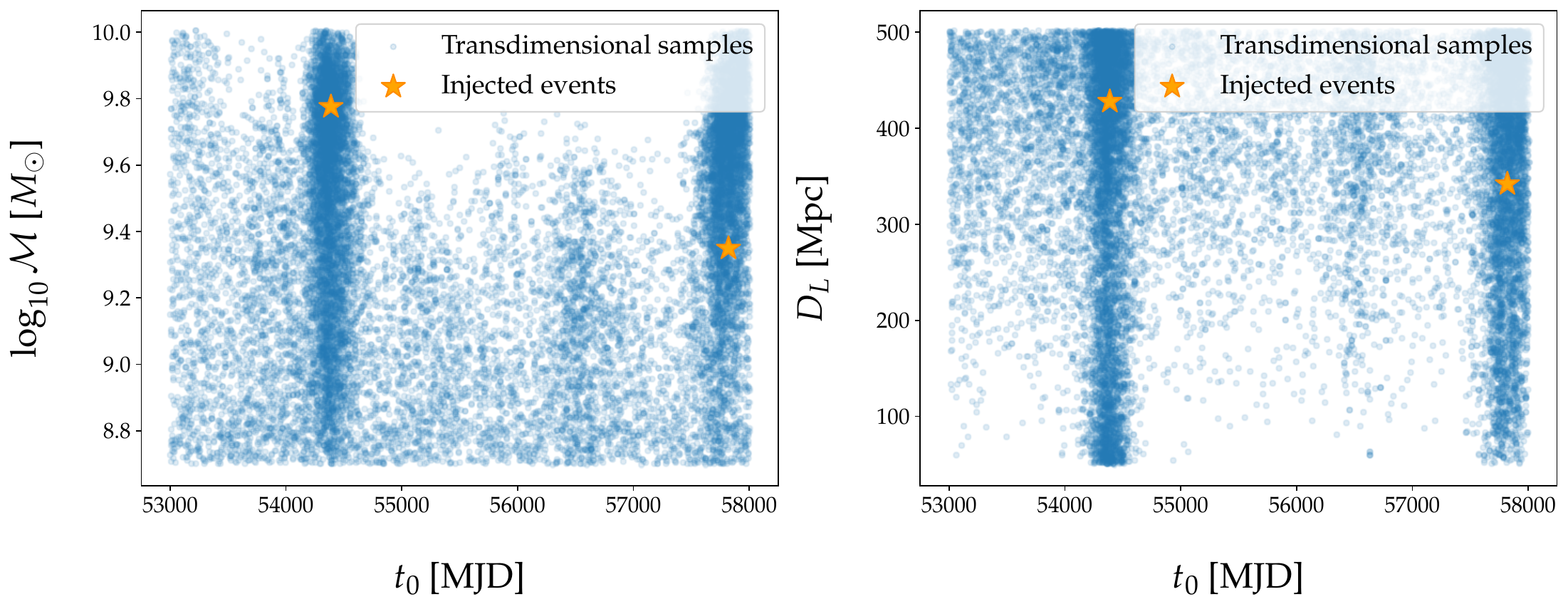}
    \caption{\justifying
    Source-level recovery for the representative toy transdimensional analysis. The blue points show posterior samples for merger signals from
    states with different catalog sizes, while the orange stars mark the two
    simulated events. The left panel shows merger time \(t_0\) versus chirp
    mass \(\log_{10}\mathcal{M}\), and the right panel shows \(t_0\) versus
    luminosity distance \(D_L\).
    \label{fig:toy_injected_recovered_app}
    }
\end{figure*}

We next examine the source-level posterior for the same realization. To visualize how the transdimensional sampler recovers the simulated merger events, Fig.~\ref{fig:toy_injected_recovered_app} compares posterior samples for the inferred merger signals with the simulated source. The blue points show the recovered source parameters for all merger signals supported by the posterior, including samples with different inferred numbers of sources, and the orange star markers show the simulated events. 

\begin{figure*}[t]
    \centering
    \includegraphics[width=0.91\linewidth]{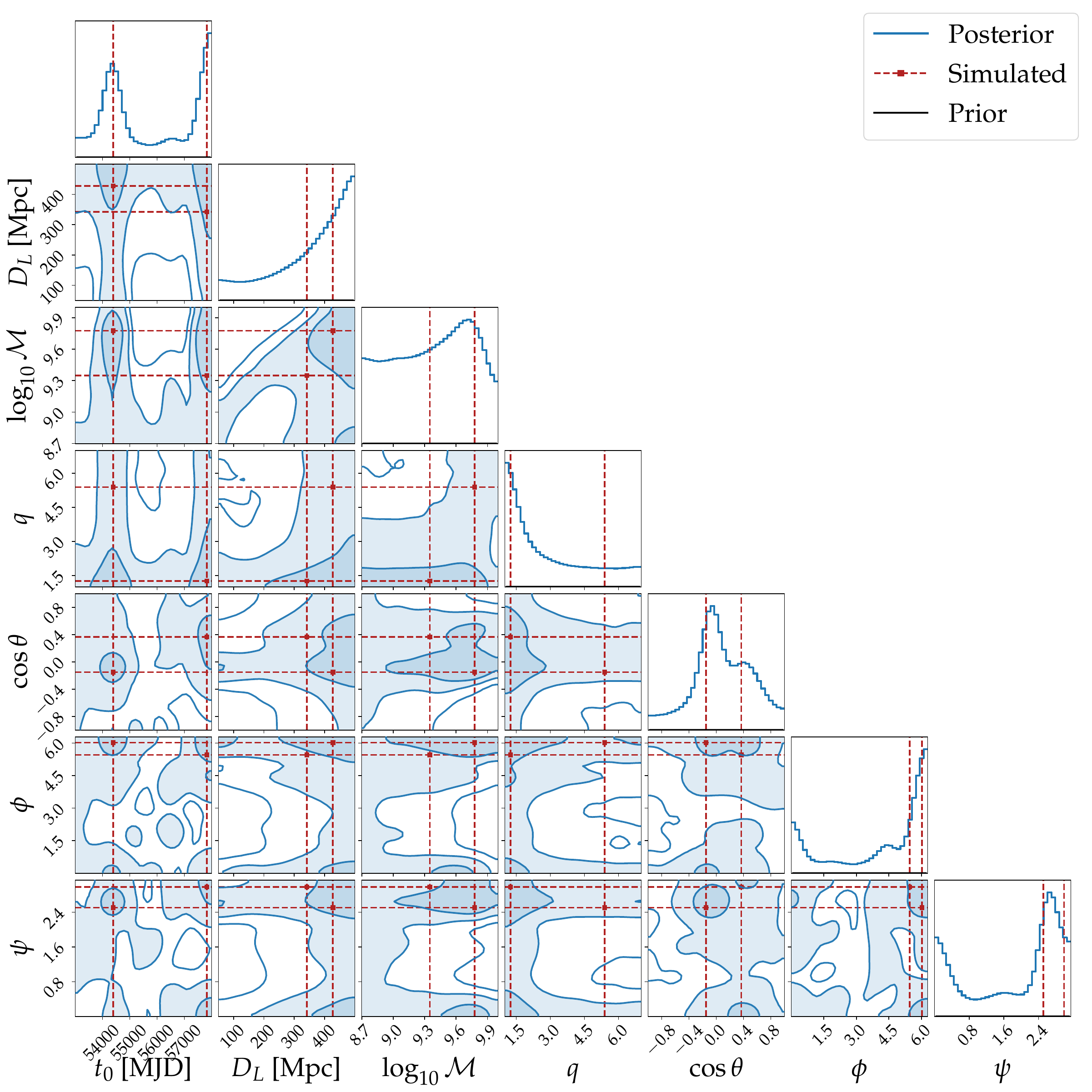}
    \caption{\justifying
    Posterior samples for the source parameters of the inferred merger signals in the representative toy RJMCMC run. Because the catalog size is not fixed, the distributions combine source samples from posterior states containing different numbers of merger signals. The blue contours show the inferred posterior samples for all merger signals, and the red dashed lines mark the simulated source parameters.}
    \label{fig:toy_source_corner_app}
\end{figure*}

Figure~\ref{fig:toy_source_corner_app} shows the joint distributions of the
source parameters for the merger signals sampled by the transdimensional
analysis. Because the catalog size varies during the RJMCMC, these
distributions pool source samples from posterior states containing different
numbers of active merger signals. The figure therefore summarizes the overall
source-parameter structure supported by the data rather than assigning a
separate posterior to each simulated event.

\begin{figure}[t] 
\centering \includegraphics[width=0.98\linewidth]{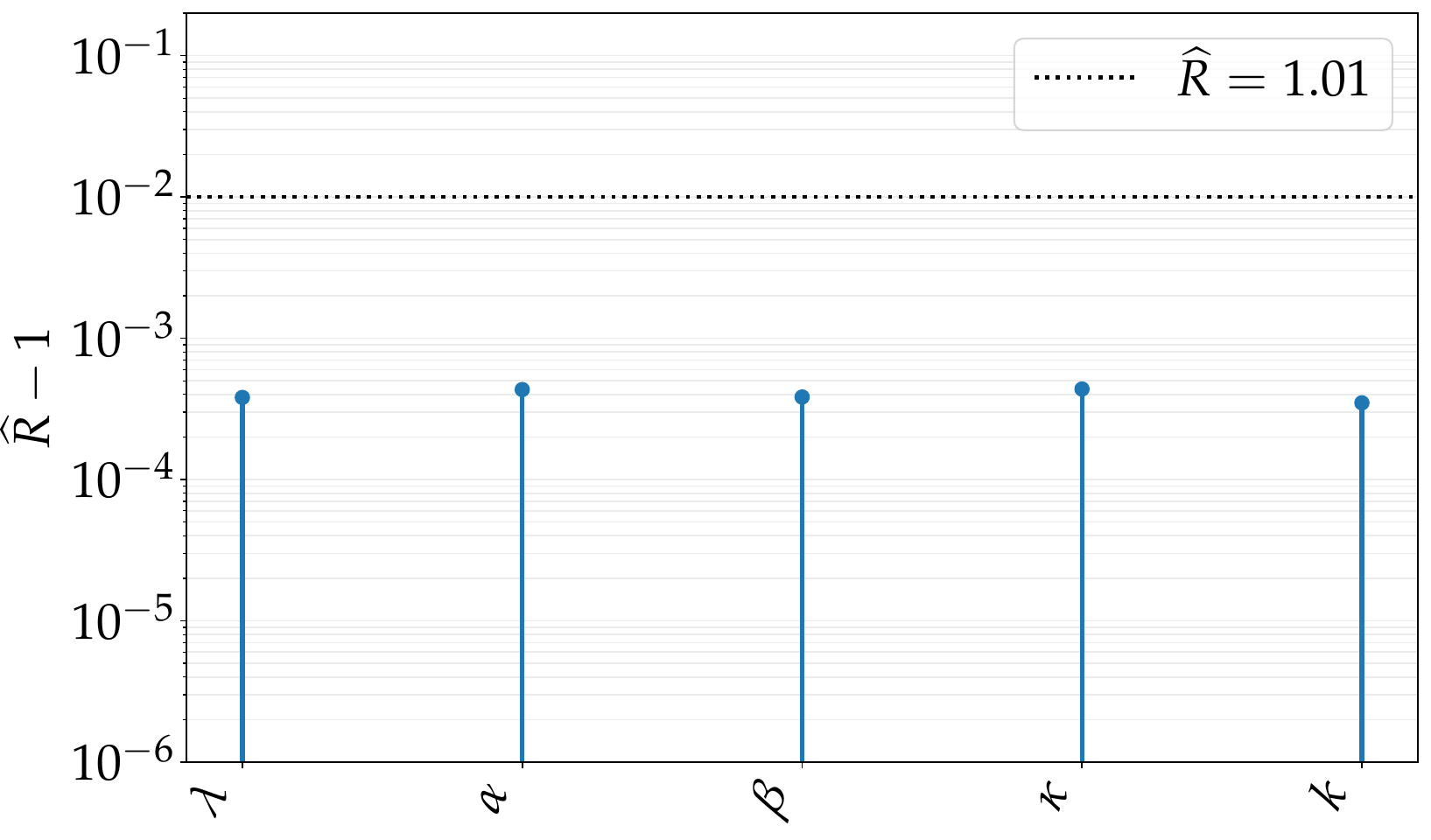} 
\caption{\justifying Convergence diagnostic for the representative toy transdimensional analysis. The vertical lines show \(\widehat{R}-1\) for the four toy-population hyperparameters \((\lambda,\alpha,\beta,\kappa)\) and the sampled catalog size \(k\). The horizontal dotted line at \(\widehat{R}-1=10^{-2}\) marks the reference value. Values closer to zero indicate closer agreement among the sampled chains. All quantities shown lie below the \(\widehat{R}=1.01\) reference threshold.} \label{fig:toy_rhat_app}
\end{figure}

Figure~\ref{fig:toy_rhat_app} shows a convergence diagnostic for the representative toy transdimensional analysis using the rank-normalized Gelman--Rubin statistic \(\widehat{R}\). We evaluate the diagnostic for the four toy-population hyperparameters \((\lambda,\alpha,\beta,\kappa)\) and for the sampled catalog size \(k\).  The horizontal dotted line at \(\widehat{R}-1=10^{-2}\) corresponds to
the reference value \(\widehat{R}=1.01\). All quantities shown lie below this reference threshold, providing an additional check that the population parameters and catalog-size sampling have converged for this run.

\bibliography{mybib,collab}

\end{document}